\documentclass[aps,prd,a4paper,superscriptaddress,nofootinbib,10pt, two column]{revtex4-1}
\usepackage{amsmath,amssymb,graphicx,multirow,dcolumn,bm,latexsym,soul,nicefrac}
\usepackage[colorlinks,linkcolor=blue,citecolor=blue,urlcolor=blue]{hyperref}
\newcounter{RomanNumber}
\usepackage[normalem]{ulem}
\newcommand{\MyRoman}[1]{\setcounter{RomanNumber}{#1}\Roman{RomanNumber}}
\allowdisplaybreaks
\newcommand{\be}{\begin{equation}}
\newcommand{\ee}{\end{equation}}
\newcommand{\bea}{\begin{eqnarray}}
\newcommand{\eea}{\end{eqnarray}}
\newcommand{\nn}{\nonumber}

\newcommand{\TRC}{TianQin Research Center for Gravitational Physics and School of Physics and Astronomy, Sun Yat-sen University (Zhuhai Campus), 2 Daxue Rd., Zhuhai 519082, P. R. China.}
\newcommand{\HUST}{MOE Key Laboratory of Fundamental Physical Quantities Measurement \& Hubei Key Laboratory of Gravitation and Quantum Physics, PGMF and School of Physics, Huazhong University of Science and Technology, Wuhan 430074, P. R. China.}
\newcommand{\UB}{School of Physics and Astronomy and Institute of Gravitational Wave Astronomy,
University of Birmingham, Edgbaston, Birmingham B15 2TT, United Kingdom}
\newcommand{\SISSA}{SISSA, Via Bonomea 265, 34136 Trieste, Italy and INFN Sezione di Trieste}
\newcommand{\IFPU}{IFPU—Institute for Fundamental Physics of the Universe, Via Beirut 2, 34014 Trieste, Italy}
\newcommand{\PA}{Institut d'Astrophysique de Paris, CNRS \& Sorbonne
 Universit\'es, UMR 7095, 98 bis bd Arago, 75014 Paris, France}

\begin{document}

\title{Science with the TianQin observatory: Preliminary results on testing the no-hair theorem with ringdown signals}

\author{Changfu Shi}
\affiliation{\TRC}
\affiliation{\HUST}

\author{Jiahui Bao}
\affiliation{\TRC}

\author{Hai-Tian Wang}

\author{Jian-dong Zhang}
\email{Emial:zhangjd9@mail.sysu.edu.cn}

\author{Yi-Ming Hu}

\affiliation{\TRC}

\author{Alberto Sesana}
\affiliation{\UB}
\author{Enrico Barausse}
\affiliation{\PA}
\affiliation{\SISSA}
\affiliation{\IFPU}

\author{Jianwei Mei}
\email{Emial:meijw@sysu.edu.cn}

\author{Jun Luo}
\affiliation{\TRC}
\affiliation{\HUST}

\date{\today}

\begin{abstract}	
We study the capability of the space-based gravitational wave observatory TianQin to test the no-hair theorem of General Relativity, using the ringdown signal from the coalescence of massive black hole binaries. We parameterize the ringdown signal by the four strongest quasinormal modes and estimate the signal to noise ratio for various source parameters. We consider  constraints  both from single detections and from all the events combined throughout the lifetime of the observatory, for different astrophysical models. We find that  at the end of the mission, TianQin will have constrained deviations  of the frequency and decay time of the dominant 22 mode from the general relativistic predictions to within 0.2 \% and 1.5 \% respectively, the frequencies of the subleading modes can be also constrained within 0.3\%. We also find that TianQin and LISA are highly complementary, by virtue of their different frequency windows. Indeed, LISA can  best perform ringdown tests for black hole masses in excess of $\sim 3\times 10^6 M_\odot$, while TianQin is best suited for lower masses.
\end{abstract}

\maketitle

\section{Introduction}

The detection of gravitational wave (GW) signals from the coalescence of compact binaries by the LIGO-Virgo-Collaboration \cite{Abbott:2016blz,Abbott:2016nmj,Abbott:2017gyy,Abbott:2017oio,Abbott:2017vtc,TheLIGOScientific:2016pea,TheLIGOScientific:2017qsa} has opened up new frontiers in testing the nature of gravity and black holes \cite{TheLIGOScientific:2016src,Berti:2018cxi,Berti:2018vdi,Alexander:2017jmt,Chamberlain:2017fjl,Yunes:2016jcc,Barausse:2016eii,Chatziioannou:2015uea}. One outstanding fact to be tested is the no-hair theorem \cite{Israel:1967wq,Israel:1967za,Carter:1971zc}, which states that within General Relativity
black holes are only characterized by mass, spin and electric charge. Since black hole electric charges
are believed to be extremely small in realistic astrophysical environments -- see e.g. \cite{Gibbons:1975kk,hanni1982limits,Goldreich:1969sb,Ruderman:1975ju,Blandford:1977ds,Barausse:2014tra} --
black holes in General Relativity  are fully determined by only two parameters, the mass and spin.

GW observations offer an experimental way to test the no-hair theorem, and thus the validity of General Relativity. After the coalescence of two  black holes, the remnant object quickly transits from a highly perturbed state to a perfect Kerr black hole, radiating a series of damped  oscillating signals (the quasinormal modes -- QNMs -- of the remnant black hole). If the no-hair theorem is correct, then the oscillation frequency and the damping time of the QNMs are completely determined by the mass and spin angular momentum of the remnant Kerr black hole. By measuring the frequency and damping time
of the least damped QNM, as well as (at least) the frequency or damping time of one of the subdominant modes, one can in principle test the no-hair theorem \cite{Dreyer:2003bv,Berti:2005ys,Berti:2007zu,Yang:2017zxs,Yang:2017xlf,Cunha:2017qtt,Barack:2018yly,Glampedakis:2017dvb,Berti:2016lat}.

Detweiler \cite{Detweiler:1980gk} first pointed out the observation of QNM may provide direct evidence of black hole.
Indeed, Dreyer et al \cite{Dreyer:2003bv} developed a formalism in which they infer the mass and angular momentum of the black hole from the strongest (least damped) QNM, and then check consistency with the no-hair theorem by considering additional subdominant modes. Berti et al \cite{Berti:2005ys,Berti:2007zu}  investigated the accuracy of parameter estimation and the capability to resolve subleading QNMs. They also provided a set of fitting functions relating the oscillation frequencies and the quality factors (a combination of frequencies and damping times)  to the mass and angular momentum of the black hole, for different QNMs. Kamaretsos et al \cite{Kamaretsos:2011um,Kamaretsos:2012bs}  introduced a set of fitting functions relating the amplitudes of several QNMs to the mass ratio and the effective spin of the progenitor binary black holes. Using this result, Gossan et al \cite{Gossan:2011ha}  investigated the capability of eLISA \cite{AmaroSeoane:2012je} and of the Einstein Telescope \cite{Punturo:2010zz} to test the no-hair theorem varying with the luminosity distance,  assuming the progenitor black holes are not spinning.

Although theoretical progress has been made, it is still difficult to test the no-hair theorem in practice. GW150914 is the first and by far the strongest GW signal from a merging black hole binary detected \cite{LIGOScientific:2018mvr} whose signal-to-noise ratio (SNR) is 24. But since the SNR for the ringdown phase is only 7 \cite{TheLIGOScientific:2016src}, so it is quite hard to extract subdominant QNM frequencies and damping times precisely for current detections\cite{Carullo:2018sfu,Carullo:2019flw,Brito:2018rfr,Isi:2019aib}. However, more accurate and precise tests will become possible with future space-based detectors and the third generation ground-based detectors~\cite{Berti:2016lat}.

In this paper, we focus on testing the no hair theorem with TianQin, a space-based GW observatory to be launched in the 2030s \cite{Luo:2015ght}.
In the center of galaxies, there exist massive black holes (MBHs) with masses ranging from $10^4 M_\odot$  to more than $10^9 M_\odot$.
The mergers of MBHs with masses in the lower part of this range ($10^4 M_\odot \sim 10^7 M_\odot$) can
be detected by TianQin with SNR above 1000 at $z \lesssim 2$ \cite{Wang:2019ryf,Feng:2019wgq,Hu:2017yoc},
and offer an excellent opportunity to test the no-hair theorem.

The paper is organized as follows.
In section \ref{sec:method}, we introduce all the necessary ingredients needed in the calculations, including the waveform for the ringdown signals, the sensitivity curve of TianQin and the statistical method.
In section \ref{sec:result}, we present the results obtained, including the SNR and the accuracy of parameter estimation for the four strongest QNMs, and the combined constraining power from all the events expected throughout the lifetime of TianQin. The combined constraining power obtained by joining the events expected from both TianQin and LISA \cite{Audley:2017drz} is also presented.
A brief summary is presented in section \ref{summary}.

Throughout this paper we set $G=c=1$.

\section{Method}\label{sec:method}

In this section, we introduce  the tools necessary for the calculations. In section \ref{subsec:signal}, we review the waveform for ringdown signals. In section \ref{subsec:detector}, we introduce the sensitivity and response of TianQin to GW signals. In section \ref{subsec:statistics}, we outline the statistical method that will be used.

\subsection{Signals}\label{subsec:signal}

We focus on the two polarizations of GWs, $h_{+,\times}\,$, as predicted by general relativity (GR). The ringdown signal from a perturbed black hole can be decomposed into a sum of QNMs, whose frequencies and decay times can be calculated by solving the linear perturbation equations with appropriate boundary conditions (outgoing at spatial infinity and ingoing at the event horizon)~\cite{Chandrasekhar:1984siy,Chandrasekhar:1975zza,Leaver:1985ax,Onozawa:1996ux,Berti:2003jh,Berti:2005eb,Kokkotas:1993ef}.
The QNMs of a Kerr black hole are conventionally labeled with three indices \((l,m,n)\),  where \(n=0,1,2,\dots\) is the overtone index, and \(l=2,3,4,\dots\) and  \(m=0,\pm 1,\dots,\pm l\) are the harmonic indices. The fundamental modes (corresponding to  \(n=0\)) usually have much larger amplitudes and much longer damping times than the higher overtones $n\geq1$ \cite{Berti:2005ys}. For this reason, in the following we will only consider the fundamental modes and denote them with two indices, $(l,m)$.

In our calculations, we will only use the ringdown part of the waveform, which  takes the form
\begin{align}
\label{waveform1}
h_{+,\times}(t)=\frac{M_z}{D_L}\sum_{l,m>0}A_{lm}Y_{+,\times}^{lm}(\iota)\Psi^{+,\times}_{lm}(t),\nn\\
\Psi^+_{lm}(t)=\exp\left(-\displaystyle\frac{t}{\tau_{lm}}\right)\cos(\omega_{lm}t-m\phi_0),\nn\\
\Psi^\times_{lm}(t)=-\exp\left(-\displaystyle\frac{t}{\tau_{lm}}\right)  \sin(\omega_{lm}t-m\phi_0),
\end{align}
for $t\geq t_0$; and $h_{+,\times}(t)=0$ for $t<t_0\,$, where $t_0$ is the starting point of the ringdown phase. \(M_z\) is the red-shifted mass of the source, \(D_L\) is the luminosity distance to the source, \(\iota\in[0,\pi]\) is the inclination angle of the source, $\phi_0$ is the initial phase, and \(A_{lm}\), \(\omega_{lm}\) and \(\tau_{lm}\) are the amplitude, the oscillation frequency and the damping time of the corresponding QNM, respectively. The functions \(Y_{\pm}^{lm}(\iota)\) can be expressed as  sums of -2 weighted spin spherical harmonics \cite{Kamaretsos:2011um}:
\begin{eqnarray}
\label{spin weighted}
Y_+^{lm}(\iota)={}_{-2}Y^{lm}(\iota,0)+(-1)^l {}_{-2}Y^{l-m}(\iota,0),\nn\\
Y_\times^{lm}(\iota)={}_{-2}Y^{lm}(\iota,0)-(-1)^l {}_{-2}Y^{l-m}(\iota,0).
\end{eqnarray}

One can approximate the ringdown signal with the combination of the few strongest QNMs. Considering the $(2,2)$, $(3,3)$, $(4,4)$ and $(2,1)$ modes, one has
\bea
\label{angular function}
&Y_+^{22}(\iota)=\displaystyle\sqrt{\frac{5}{4\pi}}\frac{1+\cos^2\iota}{2}, ~~~Y_\times^{22}(\iota)=\sqrt{\frac{5}{4\pi}}\cos \iota,\nn\\
&Y_+^{21}(\iota)=\displaystyle\sqrt{\frac{5}{4\pi}}\sin\iota,\quad Y_\times^{21}(\iota)\sqrt{\frac{5}{4\pi}}\cos\iota\sin\iota\nn\\
&Y_+^{33}(\iota)=-\displaystyle\sqrt{\frac{21}{8\pi}}\frac{1+\cos^2\iota}{2}\sin\iota,\nn\\
&Y_\times^{33}(\iota)=-\sqrt{\frac{21}{8\pi}}\sin\iota\cos\iota,\nn\\
&Y_+^{44}(\iota)=\displaystyle\sqrt{\frac{63}{16\pi}}\frac{1+\cos^2\iota}{2}\sin^2\iota,\nn\\ &Y_\times^{44}(\iota)=\sqrt{\frac{63}{16\pi}}\cos^2\iota\sin^2\iota.
\eea
By fitting numerical results, phenomenological expressions for the amplitudes have been obtained by Kamaretsos et al \cite{Kamaretsos:2012bs,Meidam:2014jpa}:
\bea
\label{amplitude}
&&A_{22}(\nu)=0.864\nu,\nn\\
&&A_{21}(\nu)=0.43[\sqrt{1-4\nu}-\chi_{eff}]A_{22}(\nu),\nn\\
&&A_{33}(\nu)=0.44(1-4\nu)^{0.45}A_{22}(\nu),\nn\\
&&A_{44}(\nu)=[5.4(\nu-0.22)^2+0.04]A_{22}(\nu),
\eea
where \(\nu=m_1m_2/(m_1+m_2)^2\) is the symmetric mass ratio, and
\bea
\label{efficient spin}
\chi_{eff}=\frac{1}{2}\Big(\sqrt{1-4\nu}\chi_1+\frac{m_1\chi_1-m_2\chi_2}{m_1+m_2}\Big)\,.
\eea
Here \((m_1,m_2)\) and \((\chi_1,\chi_2)\) are the masses and spin parameters of the progenitor black holes,  $\chi_i=J_i/m_i^2$, where $J_i$ are the spin angular momentum as we assume aligned binaries.

It should be noted that there already exist waveform models for the ringdown with consideration of relative phases between different modes \cite{London:2014cma,London:2018gaq}, and inspiral-merger-ringdown waveforms with multiple modes calibrated to numerical relativity simulations \cite{Brito:2018rfr}. Although a more accurate waveform model is essential to perform a robust parameter estimation with real data, for this work whose main objective is to obtain the projected precision, the model employed here is sufficient.

Berti et al \cite{Berti:2005ys} have provided fitting formulae relating the oscillation frequencies and damping times to the mass \(M_z\) and spin parameter $\chi_f$ of the remnant Kerr black hole:
\bea
\label{fitting formula}
\omega_{GR}&=&\frac{f_1+f_2(1-\chi_f)^{f_3}}{M_z},\nn\\
\tau_{GR}&=&\frac{2(q_1+q_2(1-\chi_f)^{q_3})}{\omega_{GR}}\,,
\eea
where the coefficients are listed in Table \ref{tablefc}.

\begin{table*}[!htbp]
		\caption{Fitting coefficients for equation \eqref{fitting formula}. Taken from \cite{Berti:2005ys}.}
		\label{tablefc}	\begin{tabular}{|p{1.5cm}<{\centering}|p{1.5cm}<{\centering}|p{1.5cm}<{\centering}|p{1.5cm}<{\centering}|p{1.5cm}<{\centering}|p{1.5cm}<{\centering}|p{1.5cm}<{\centering}|}
		\hline
		\cline{1-7}
		\((l,m)\)&\(f_1\)&\(f_2\)&\(f_3\)&\(q_1\)&\(q_2\)&\(q_3\)\\
		\cline{1-7}
		\((2,1)\)&0.6000&-0.2339&0.4175&-0.3000&2.3561&-0.2277\\
		\cline{1-7}
		\((2,2)\)&1.5251&-1.1568&0.1292&0.7000&1.4187&-0.4990\\
		\cline{1-7}
		\((3,3)\)&1.8956&-1.3043&0.1818&0.9000&2.3430&-0.4810\\
		\cline{1-7}
		\((4,4)\)&2.3000&-1.5056&0.2244&1.1929&3.1191&-0.4825\\
		\hline
	\end{tabular}
\end{table*}

The final mass $M_z$ and spin parameter $\chi_f$ of the remnant black hole
can be computed from the parameters \((m_1,m_2,\chi_1,\chi_2)\) of
the progenitor binary, e.g. via the formulae of \cite{Husa:2015iqa,Hofmann:2016yih,Barausse:2012qz}, which reproduce the results
of numerical relativity simulations.

Following \cite{Li:2011cg,Gossan:2011ha}, we parametrize possible deviations from GR with a set of dimensionless parameters, \((\delta\omega_{lm}, \delta\tau_{lm})\),
which we assume to be independent of the other source parameters and which are
defined via
\bea
\label{nonGR}
&\omega_{lm}=\omega_{lm,GR}(1+\delta\omega_{lm}),\\
&\tau_{lm}=\tau_{lm,GR}(1+\delta\tau_{lm}),
\eea
where \(\omega_{lm,GR}\) and \(\tau_{lm,GR}\) denote the GR predictions.
A violation of the no-hair theorem corresponds to at least one of the $\delta\omega_{lm}$'s and $\delta\tau_{lm}$'s being non-zero.

\subsection{Detector response}\label{subsec:detector}

Our main objective is to evaluate the scientific performance of TianQin \cite{Luo:2015ght}, focusing on tests of the no-hair theorem with ringdown signals.

TianQin will consist of a constellation of three satellites on a geocentric orbit with radius of about $10^5$ km. The three satellites are spaced evenly on the orbit to form a nearly equilateral triangle. Test masses are carried by the satellites, which are drag-free controlled to suppress non-gravitational disturbances, so that the test masses can
move along geodesics as much as possible. Laser interferometry
between the test masses is then used to detect GWs.

TianQin adopts a ``3 month on + 3 month off'' observation scheme to cope with the thermal problem faced by geocentric GW missions. It will be interesting to consider a scenario
in which twin constellations of TianQin  are present,
 with orbital planes perpendicular to each other and both nearly perpendicular to the ecliptic. In this case, the twin constellations can operate in alternation to fill up the observation gaps. Note that this scheme will not affect the sensitivity of each detector.

We adopt the following model for noise of TianQin \cite{Luo:2015ght}:
\begin{align}
\label{noiseTianQin}
S_N(f)=\frac{4S_a}{(2\pi f)^4L_0^2}\Big(1+\frac{10^{-4}{\rm Hz}}{f}\Big) +\frac{S_x(f)}{L_0^2}
\end{align}
where \(L_0=\sqrt{3}\times 10^8{\rm m}\), \(\sqrt{S_a}=1\times 10^{-15}{\rm ms^{-2}Hz^{-1/2}}\) is the average residual acceleration on a test mass, and \(\sqrt{S_x}=1\times10^{-12}~{\rm mHz^{-1/2}}\) is the total displacement noise in a single link.
Then, the sky averaged sensitivity of TianQin can be modeled by \cite{Luo:2015ght,Hu:2018yqb,Wang:2019ryf}:
\begin{align}
\label{curveTianQin}
&S_n^{SA}(f)=\frac{S_N(f)}{\bar{R}(f)},\nn\\
&\bar{R}(f)\simeq\frac{3}{10}\Big[1+\Big(\frac{2fL_0}{0.41c}\Big)^2\Big]^{-1},
\end{align}
where  $c$ is the speed of light, $\bar{R}$ is the sky averaged response function (the factor of $3/10$ comes from the angle between the detector arms, the averaged of sky location angle  of the source and the polarization angle, and the contribution of two independent Michelson's interferometers). We adopt a conservative lower frequency cutoff at $10^{-4} \rm Hz$, $i.e.$ the upper bound of the red-shift mass corresponds about $10^8 M_\odot$,  for the low frequency behavior of the Tianqin acceleration noise is not clear at this moment \cite{Wang:2019ryf}.

\begin{figure}[htbp]
	\centering
	\includegraphics[width=0.48\textwidth]{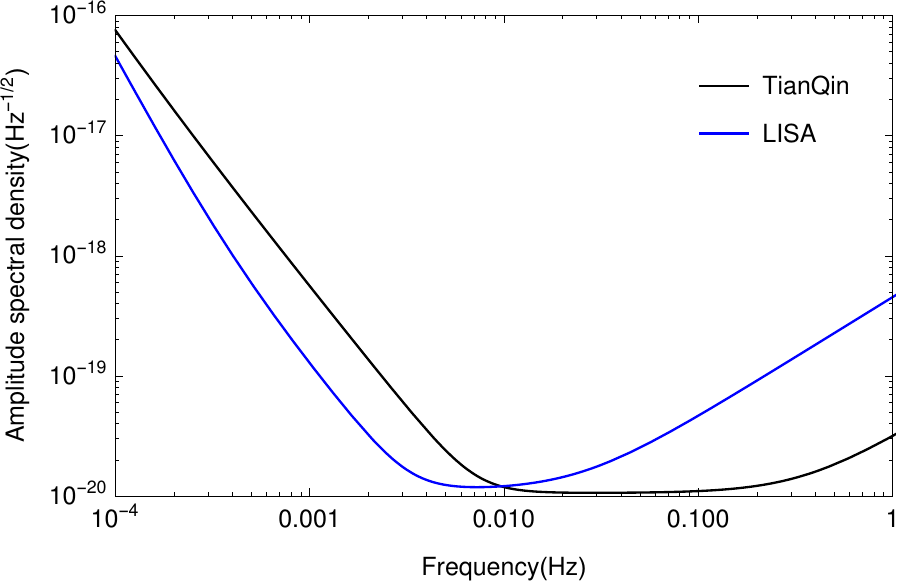}
	\caption{Anticipated sensitivity curve of TianQin and LISA.}\label{fig:SC}
\end{figure}

For completeness, we will also consider a second detector, LISA \cite{Audley:2017drz}, now adopted by the European Space Agency and due to launch in the early 2030s'. We will consider the joint capability of TianQin and LISA to test the no-hair theorem. For LISA, we will use the sensitivity curve given in \cite{Cornish:2018dyw}, and the frequency bound of $10^{-4} \rm Hz$ is no longer the case. An illustration of sensitivity curve of TianQin and LISA is given in Fig:\ref{fig:SC}.

\subsection{Statistical methods}\label{subsec:statistics}
For a pair of frequency domain signals  \(p(f)\) and \(q(f)\), one can define the inner product \cite{Finn:1992wt}.
\bea
\label{inner product}
(p|q)=2\int_{f_{\rm low}}^{f_{\rm high}} \frac{p^*(f)q(f)+p(f) q^*(f)}{S_n^{SA}(f)}df,
\eea
where the factor of 2 comes from the single side integration of the frequency. In order to prevent  spurious power arising from the Fourier transformation, $f_{\rm low}$ is taken to be half of the oscillation frequency for \((2,1)\) mode, and $f_{\rm high}$ is taken to be twice of the oscillation frequency for \((4,4)\) mode, which follows the choice of \cite{Gossan:2011ha}.
While our results are sensisitve to the choice of $f_{\rm low}$, choosing this lower limit of integration is analogous to setting a starting frequency for the ringdown, which is known to be a delicate problem\cite{Baibhav:2017jhs,Isi:2019aib}.
Our results are insensitive to the choice of $f_{\rm high}$, provided that it is taken to be sufficiently large.

Signals in the frequency domain are obtained from the time domain signals through the Fourier transformation:
\bea
\label{Fouries}
h(f)=\int_{-\infty}^{+\infty}h(t)\exp^{-2\pi i f t}dt\,,
\eea
and the SNR for a GW signal is simply defined as
\bea
\label{SNR}
{\rm SNR}[h]=\rho[h]=\sqrt{(h|h)}\,.
\eea
If the sources are isotropically distributed, one can get rid of the SNR dependence
on the sky position by performing an angular average. The sky averaged SNR can be given by Eq.~\eqref{SNR} with $h$ to be the signal before detector response.

In the case of large SNR signals, the uncertainty in the parameter estimation is given by
\bea
\label{estimation}
\Delta\vartheta^a\equiv\sqrt{\langle \delta\vartheta^a \delta\vartheta^a\rangle}\approx\sqrt{(\Gamma^{-1})^{aa}},
\eea
where \(\vartheta^a\) are the waveform parameters to be estimated, \(\langle\dots\rangle\) denotes the expectation value, and $\Gamma^{-1}$ is the inverse of the  Fisher information matrix \cite{Finn:1992wt,Cutler:1994ys,Vallisneri:2007ev},
\bea
\label{FIM}
\Gamma_{ab}=\Big(\frac{\partial h}{\partial \vartheta^a}\Big|\frac{\partial h}{\partial \vartheta^b}\Big)\,.
\eea

For the single ringdown signal, we deal with a 16-dimensional parameter space
\bea
\label{parameter}
\vec{\vartheta}=\{M_z,\chi_f,\chi_{eff},\nu,D_L,\iota,t_0,\phi_0,\delta\omega_{lm}, \delta\tau_{lm}\}.
\eea

Under the assumption that the no-hair theorem violating parameters $\delta\omega_{lm}$ and $\delta\tau_{lm}$ are independent of the source, more stringent constrains can be reached by combining multiple detections.
The incoherent superposition of multiple signals can be expressed as
\bea
\label{whole waveform}
h^{tot}(t)=\sum h(t).
\eea
Each single signal can be regarded as independent,
\bea
\Big(\frac{\partial h^{tot}}{\partial \vartheta^a_{i}}\Big|\frac{\partial h^{tot}}{\partial \vartheta^b_{j}}\Big) =\delta_{i,j}\Big(\frac{\partial h^{tot}}{\partial \vartheta^a_{i}}\Big|\frac{\partial h^{tot}}{\partial \vartheta^b_{i}}\Big),
\eea
with $\vec{\vartheta}_{i}=\{M_{zi},\chi_{fi},\chi_{effi},\nu_i,D_{Li},\iota_i,t_{0i},\phi_{0i}\}$ denoting the parameters related to the source $i$. In this case, the full parameter space is
\begin{align}
\label{whole parameter}
\vec{\vartheta}=\{\vec{\vartheta_{i}}, \delta\omega_{lm}, \delta\tau_{lm}\}\,.
\end{align}

\section{Results}\label{sec:result}

In this section, we present  preliminary results on TianQin-based projected tests of the no-hair theorem with ringdown signals.  In \ref{subsec:single}, we discuss the case of individual detections by TianQin alone, and by TianQin and LISA. In \ref{subsec:multiple}, we consider the case when all detected events are combined together, considering different possible detector configurations.

\subsection{Single detections}\label{subsec:single}

We start here by considering individual detections. For simplicity, we consider a mildly inclined binary with \(\iota=\pi/3\), and without loss of generality we set \(t_0=0\) and \(\phi_0=0\).

\begin{figure}[htbp]
\begin{minipage}{0.5\textwidth}
\includegraphics[width=0.45\textwidth]{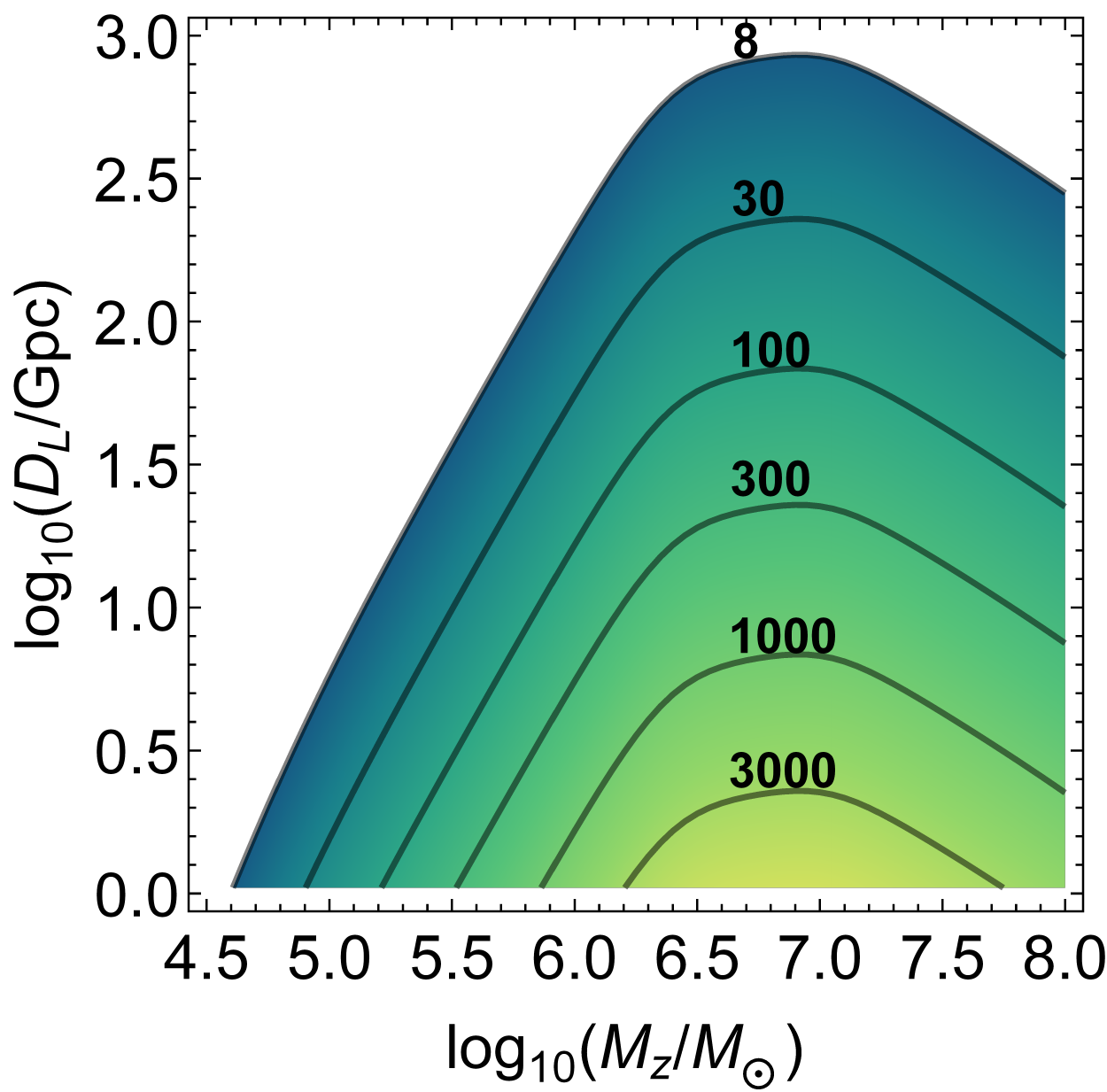}
\includegraphics[width=0.45\textwidth]{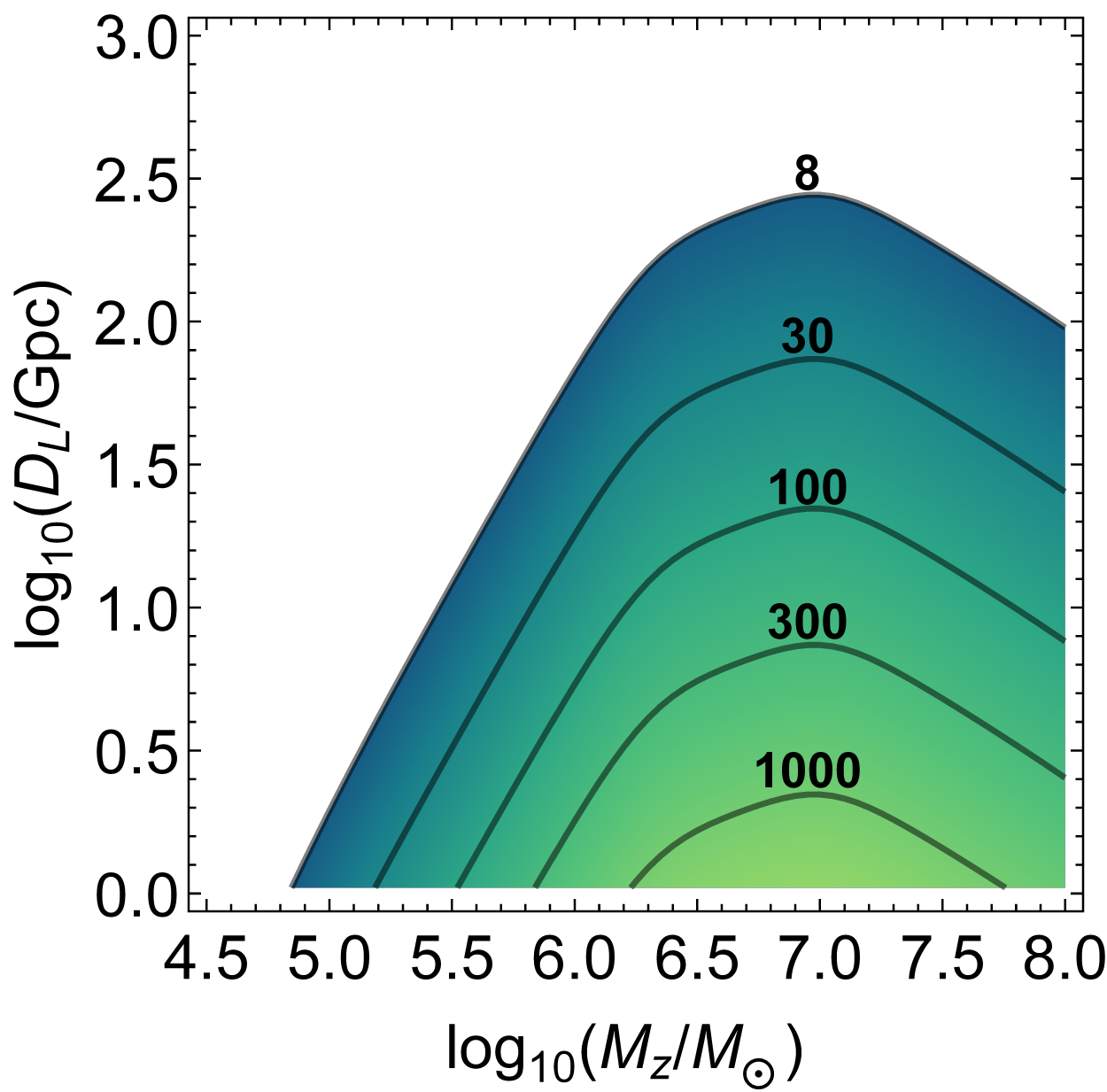}\\
\includegraphics[width=0.45\textwidth]{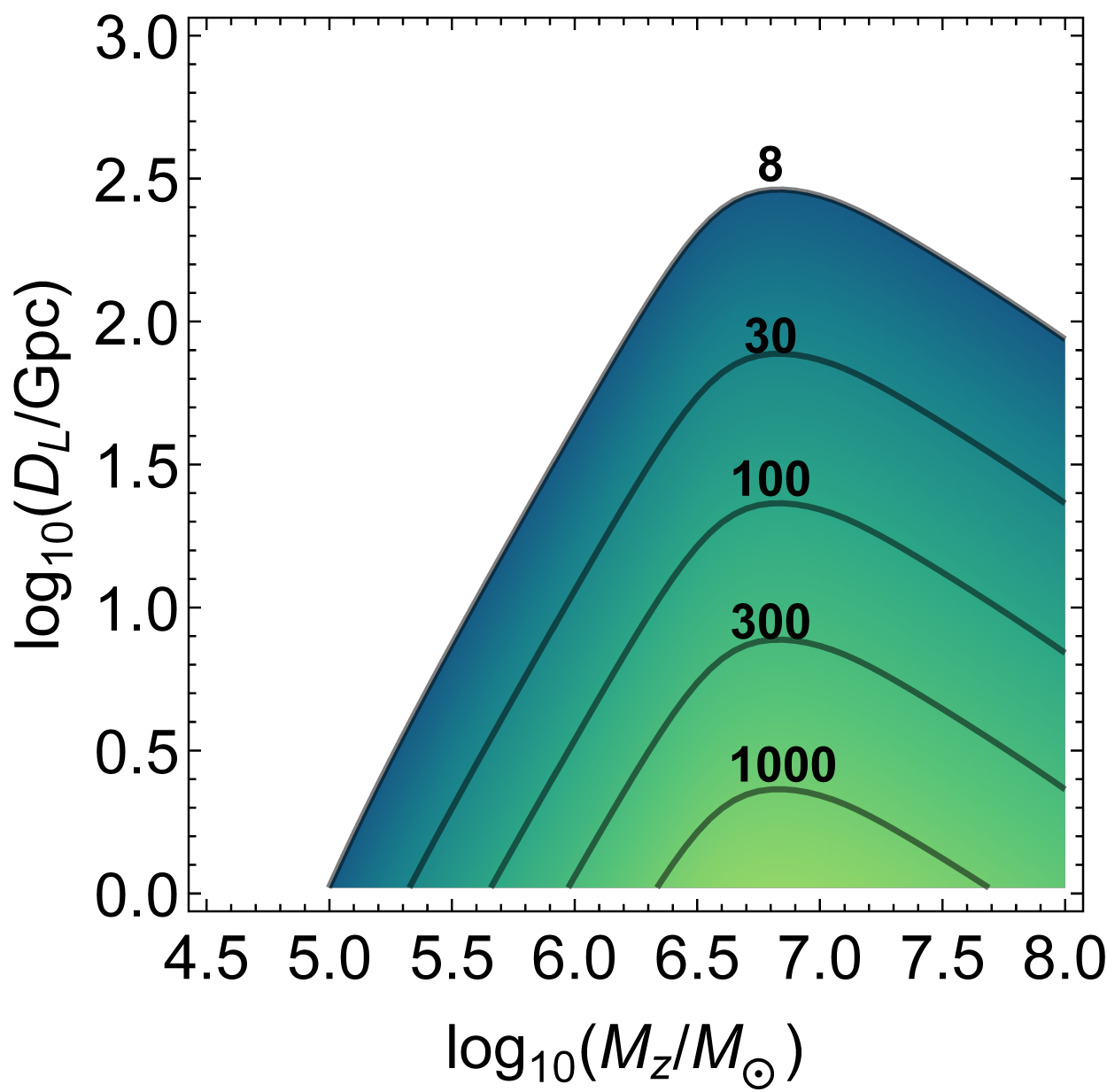}
\includegraphics[width=0.45\textwidth]{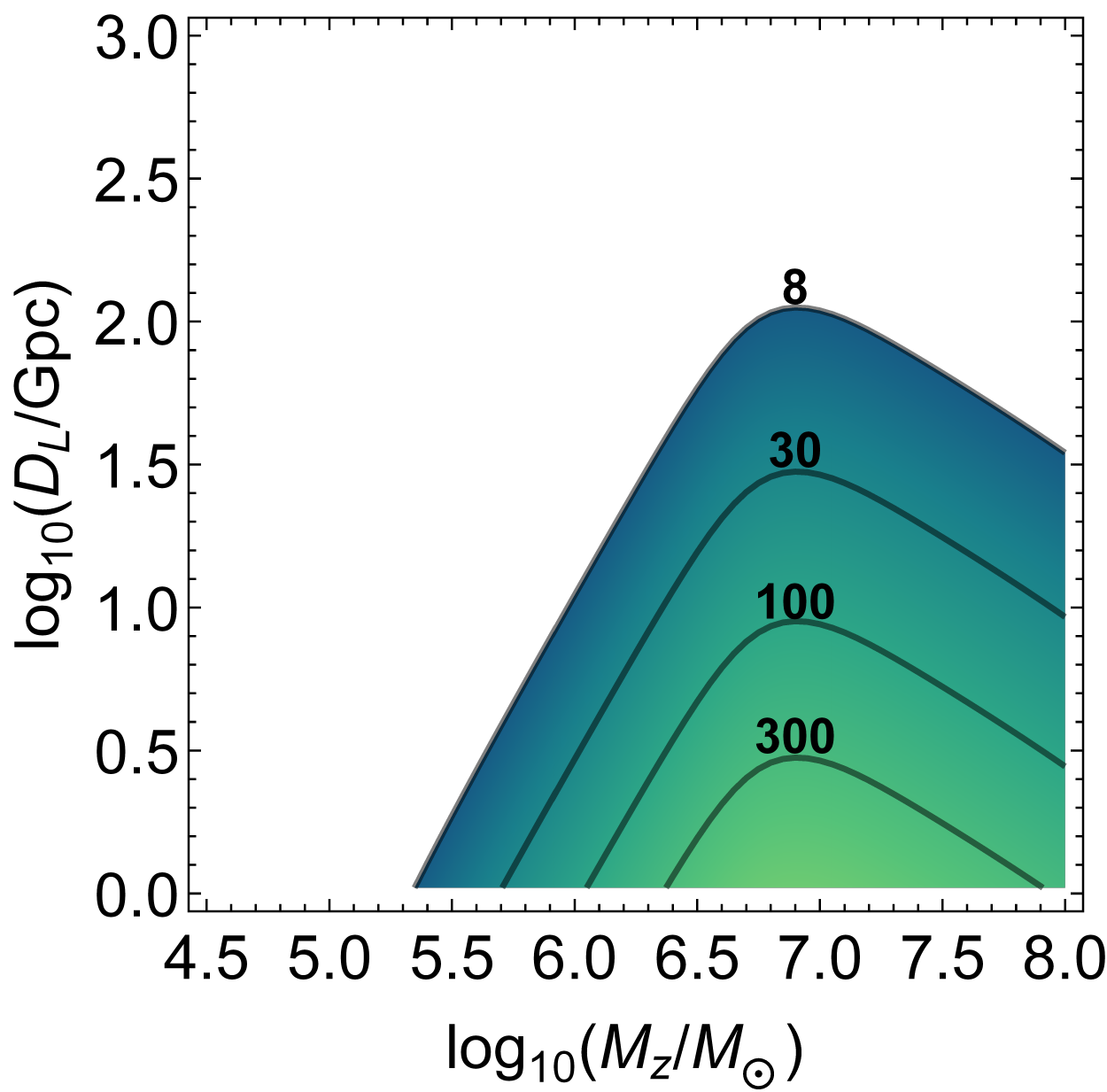}\\
\includegraphics[width=0.5\textwidth]{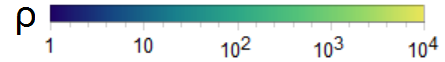}
	\caption{SNR of the four strongest QNMs in the ringdown signal for an event with \(\chi_f=0.76\), \(\nu=2/9\), \(\chi_{eff}=-0.3\). The top left figure is for the 22 mode, the top right figure is for the 21 mode, the bottom left figure is for the 33 mode, and the bottom right figure is for the 44 mode.}\label{figSNR}
	\end{minipage}
\end{figure}

Contour plots of the SNR of the four strongest QNMs as a function of the red-shifted final mass and luminosity distance are given in Fig. \ref{figSNR}. One can see that the \((2,2)\) mode is the strongest, while the \((2,1)\) and the \((3,3)\) modes are comparable to one another.

\begin{figure}[htbp]
	\centering
	\includegraphics[width=0.49\textwidth]{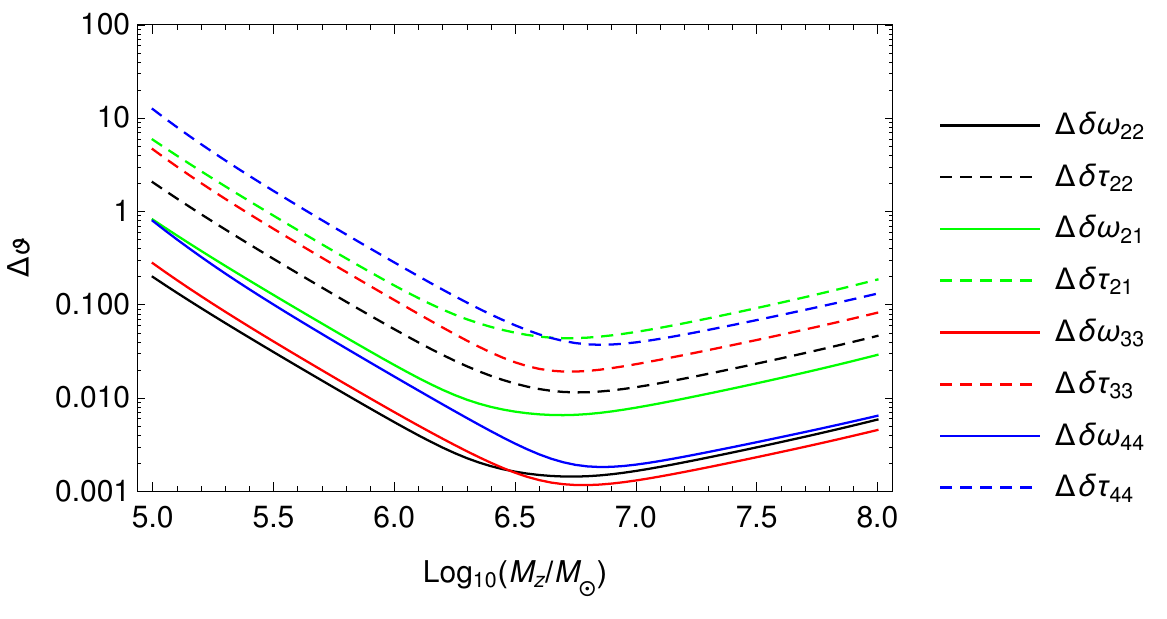}
	\caption{Parameter estimation accuracy as a function of the observed final black hole mass. Other parameters used for this plot are \(D_L=15\rm Gpc\) (i.e. red-shift \(z\approx2\) for \(\Lambda\)CDM model), \(\chi_f=0.76, \nu=2/9, \chi_{eff}=-0.3\).}\label{fig:Allmode1}
\end{figure}

The dependence of the parameter estimation accuracy  on the black hole mass is illustrated in Fig. \ref{fig:Allmode1}. The observed total mass affects the parameter estimation in two ways, i.e. through the GW amplitude and the GW frequencies. For \(M_z\lesssim3\times10^6 \rm M_\odot\), the amplitude effect dominates and the constraints get worse with smaller mass
(since the amplitude scales linearly with the mass). For \(M_z\gtrsim3\times10^6 \rm M_\odot\), the frequency effect dominates, and the constraints get worse with larger mass (because the latter leads to lower GW frequencies, which eventually fall outside the most sensitive frequency band of TianQin).

\begin{figure}[htbp]
	\centering
	\includegraphics[width=0.49\textwidth]{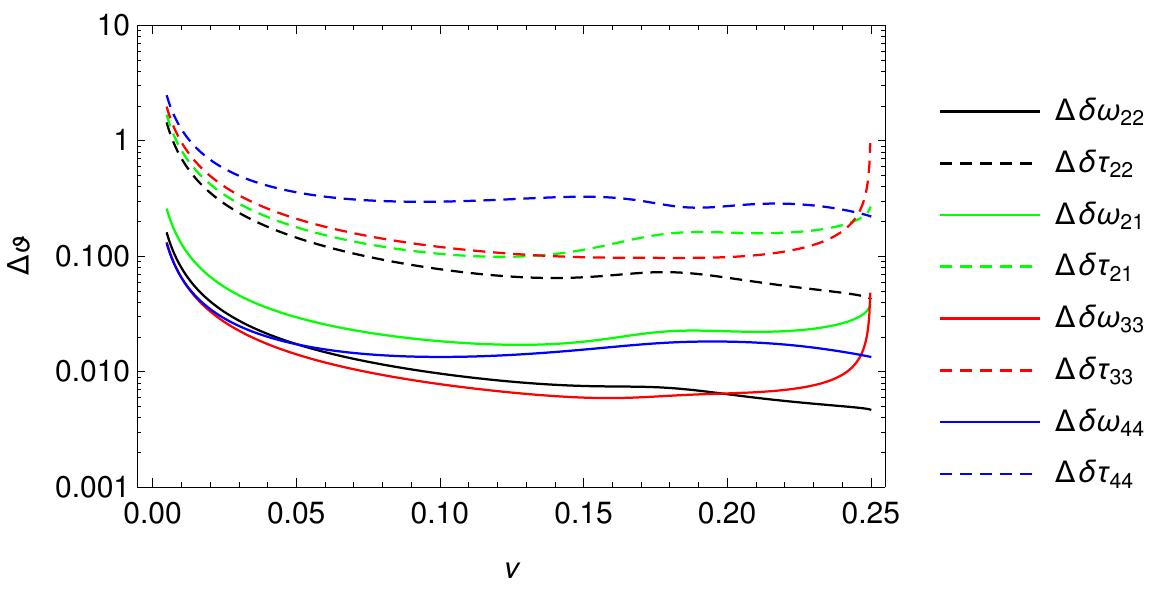}
	\caption{Parameter estimation accuracy vs. symmetric mass ratio $\nu$. Other parameters used for this plot are \(M_z=10^6 \rm M_\odot, D_L=15 \rm Gpc,\)\( \chi_f=0.76, \chi_{eff}=-0.3\).}\label{fig:Allmode2}
\end{figure}

Fig. \ref{fig:Allmode2} illustrates the dependence of the parameter estimation accuracy  on the symmetric mass ratio of the progenitor binary. The first feature is that the accuracy becomes worse for smaller mass ratios. This is because the radiated energy becomes smaller with smaller symmetric mass ratio, when the total mass is fixed. The second feature is that the constraint on the \((3,3)\) mode becomes worse as the binary masses become comparable. This is because the amplitude of the \((3,3)\) mode becomes zero when \(\nu\rightarrow 1/4\), as it is obvious from equation \eqref{amplitude}. In that limit, one should use other parameters, such as \(\delta\omega_{21}\) and \(\delta\omega_{44}\), to test the no-hair theorem.

\begin{figure}[htbp]
	\centering
	\includegraphics[width=0.49\textwidth]{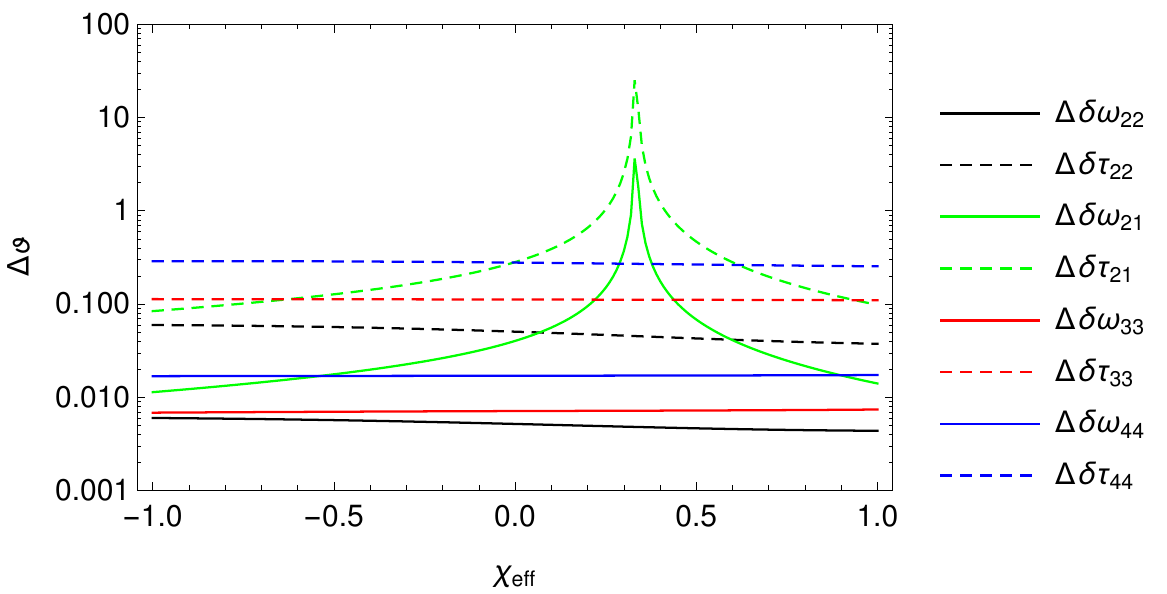}
	\caption{Parameter estimation accuracy versus effective spin $\chi_{eff}$. Other parameters used for this plot are: \(M_z=10^6 \rm M_\odot, D_L=15 \rm Gpc,\)\( \chi_f=0.76, \nu=2/9\).}\label{fig:Allmode3}
\end{figure}

The dependence of the parameter estimation accuracy on the effective spin is plotted in Fig. \ref{fig:Allmode3}. The effective spin has negligible effect on parameter constraints, except for $\delta\omega_{21}$ and $\delta\tau_{21}$. The divergence of $\Delta\delta\omega_{21}$ and $\Delta\delta\tau_{21}$ appears because the amplitude of the 21 mode tends to zero when $\chi_{eff}=\sqrt{1-4\nu}$, as shown in equation (\ref{amplitude})

\begin{figure}[htbp]
	\centering
	\includegraphics[width=0.49\textwidth]{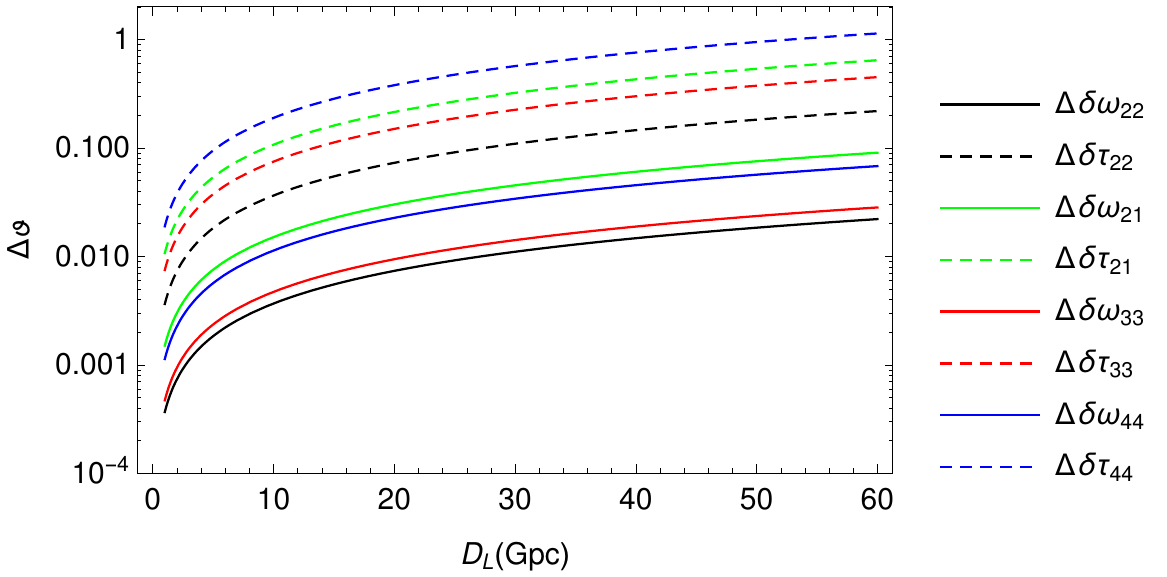}
	\caption{Parameter estimation accuracy vs. luminosity distance. Other parameters for this plot are  \(M_z=10^6 \rm M_\odot,\)\( \chi_f=0.76, \chi_{eff}=-0.3, \nu=2/9\).}\label{figdl}
\end{figure}

The dependence of the parameter estimation accuracy on the luminosity distance is shown in Fig. \ref{figdl}. The errors are inversely proportional to $D_L$. This is because the amplitude of the GW is proportional to $1/D_L$, and as a result all components of the covariance matrix are proportional to to $1/D_L^2$.

\begin{figure}[htbp]
	\centering
	\includegraphics[width=0.49\textwidth]{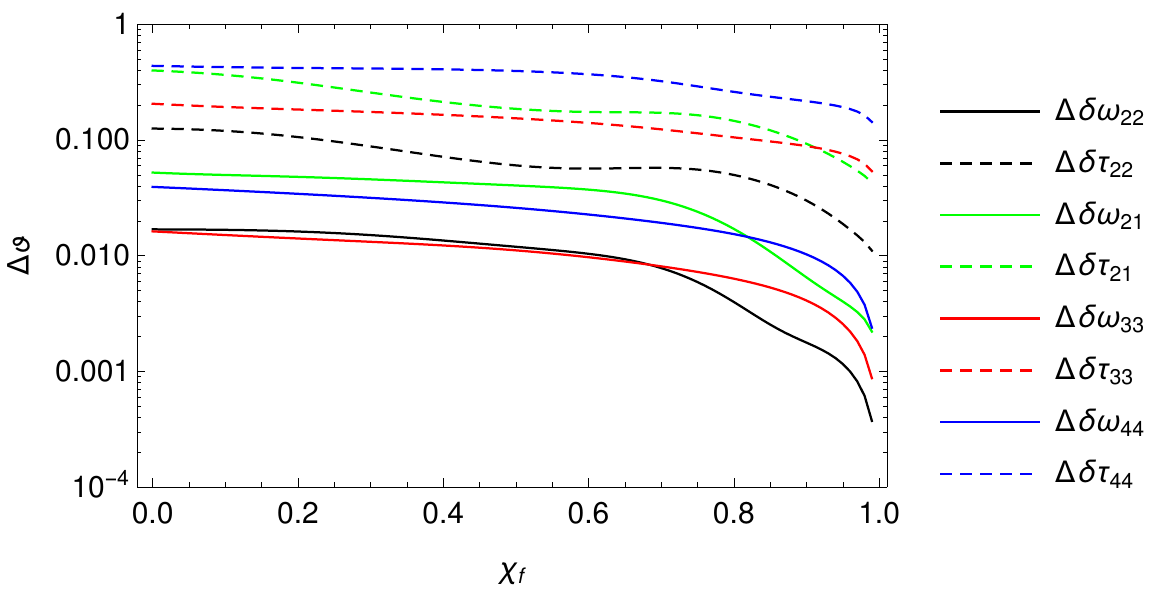}
	\caption{Parameter estimation accuracy vs. spin of the final black hole. Other parameters for this plot are  \(M_z=10^6 \rm M_\odot, \chi_{eff}=-0.3, \nu=2/9, D_L=15Gpc\).}\label{figsp}
\end{figure}

Fig. \ref{figsp} shows the parameter estimation accuracy as a function of the final spin. The radiated GW frequencies become larger for larger spins, and so is the constraining power.

From the above figures, one can see that \(\delta\omega_{22}\), \(\delta\tau_{22}\) and \(\delta\omega_{33}\) are typically the most constrained parameters. We thus will focus on these parameters in the following.

Since LISA and TianQin should be launched around the same time, we can also consider a joint detection by the two missions.  The two detectors are most sensitive at different frequencies, and the frequency of the ringdown signal is inversely proportional to the mass of the remnant. It is therefore interesting to investigate the capabilities of the two instruments (and of the combined observations) as a function of remnant black hole mass.

\begin{figure}[htbp]
	\centering
	\includegraphics[width=0.45\textwidth]{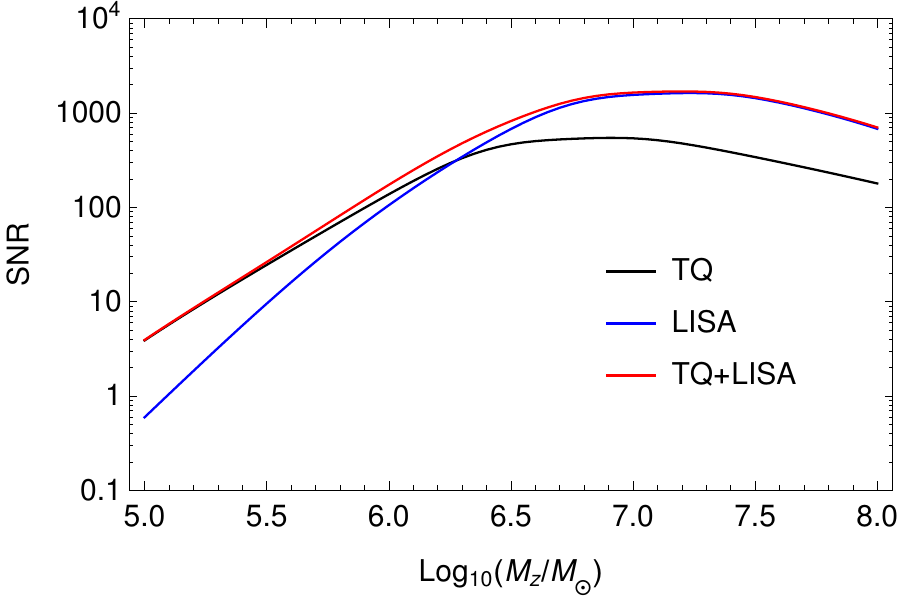}
	\caption{Ringdown SNR as a function of the observed black hole mass for TianQin, LISA and for a joint detection. Other parameters used for this plot are \(\chi_f=0.76,\rm, \chi_{eff}=-0.3, \nu=2/9,D_L=15Gpc. \)}\label{figSNR1}
\end{figure}

An illustration of the ringdown SNR for TianQin, LISA and a joint detection as a function of the observed final mass is given in Fig. \ref{figSNR1}. Around \(M_z\sim 10^6M_\odot\), when the SNRs for both detectors are comparable, a joint detection can improve the total SNR by a factor of about 1.4 at best.

\begin{figure}[htbp]
	\centering
	\includegraphics[width=0.50\textwidth]{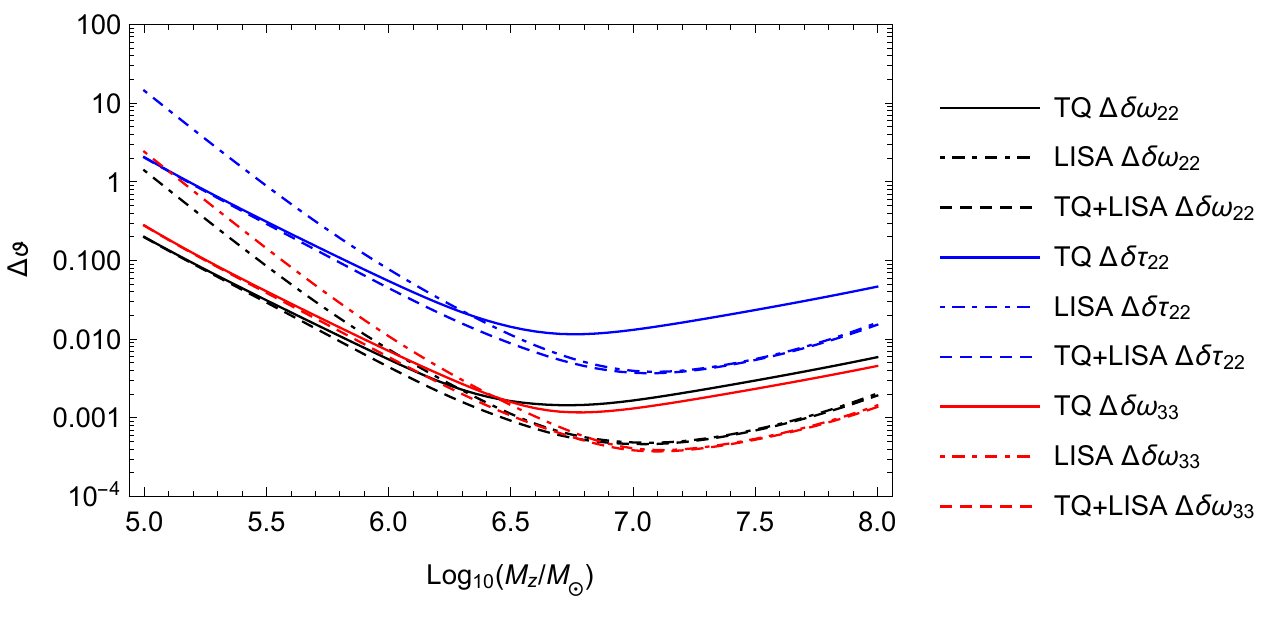}
	\caption{\(\delta\omega_{22}\) (black),  \(\delta\tau_{22}\) (blue) and \(\delta\omega_{33}\) (red) estimation accuracy for TianQin (full line), LISA (dot-dash line) and a joint detection (dash line), as a function of the observed final black hole mass. Other parameters for this plot are \(\chi_f=0.76,\rm, \chi_{eff}=-0.3, \nu=2/9,D_L=15Gpc.\)}\label{figmass}
\end{figure}

Fig. \ref{figmass} shows the estimation accuracy of \(\delta\omega_{22}\) (black),  \(\delta\tau_{22}\) (blue) and \(\delta\omega_{33}\) (red) estimation accuracy for TianQin (full line), LISA (dot-dash line) and a joint detection (dash line), as a function of the observed final mass. The constraint can be improved by a factor of about 1.4 near \(M_z\sim10^6\rm M_\odot\) with a joint detection.

This nicely shows the complementarity of LISA and TianQin. The former will dominate the measurement for \(M_z>3\times 10^6\rm M_\odot\) while the latter will provide much better constraints for \(M_z<10^6\rm M_\odot\).

\subsection{Combined constrains from all observed events}\label{subsec:multiple}

Massive black hole population models predict that TianQin can detect from tens to few hundreds of massive black hole mergers during its five years operation time \cite{Wang:2019ryf,Feng:2019wgq}. In this subsection, we study how TianQin and LISA can test the no-hair theorem by combining their massive black hole detections. As a comparison, we will also consider the case for both sigle constellation and a  twin constellation of TianQin running for 5 years (labeled as ``TQ" and ``TQ\_tc" respectively), LISA running for 4 years (labeled as ``LISA\_4y") and LISA running for 10 years (labeled as ``LISA\_10y").

The number and properties of massive black hole mergers that can be detected are largely model dependent. We will use the same three scenarios for the merger history of massive black holes investigated in \cite{Klein16,Wang:2019ryf} and generated according to the semi-analytic model presented in \cite{EB12} and successively improved in \cite{Sesana14,Antonini_long}.
The three scenarios are referred to as ``popIII", ``Q3\_d" and ``Q3\_nod", corresponding, respectively, to a light seed model \cite{Madau:2001sc}  and to two heavy seed models \cite{Bromm:2002hb,Begelman:2006db,Lodato:2006hw} with and without time delays between the mergers of massive black holes and those of their host galaxies. We refer the readers to \cite{EB12,Sesana14,Antonini_long} for a detailed description of these three models.

For each of the detector scenarios, we produce 1000 mock catalogues of observed events from each of the three astrophysical models.
Each catalogue consists of all the events that can be detected under the corresponding detector scenario. We assume a detection threshold of 8 for the SNR of total Inspiral-Merger-Ringdown(IMR) signal ($\rho_{\rm IMR}>8$),  we also assume a detection threshold of 8 for the SNR of total Ringdown stage ($\rho_{\rm Rd}>8$). For a given detector scenario, the expected IMR detection number and Ringdown detection number are obtained by averaging over the 1000 mock catalogues.

Using the criterion provided in \cite{Berti:2016lat,Berti:2007zu} (see in particular
Eqs. 2 and 3 of \cite{Berti:2016lat} for readily usable formulae), $i.e.$ $\rho_{\rm Rd}>\rho_{\rm GLRT}$, where \(\rho_{\rm GLRT}\) is the SNR for generalized likelihood ratio test, \(i.e.\) requiring
that the SNR in the ringdown alone should be sufficient for
the first subleading mode to be resolvable from the leading one,  one can calculate the number of events that can be used to test the no-hair theorem. We call these events the ``testing events''. For a given detector scenario, the expected number of
testing events is also obtained by averaging over the mock catalogues.

\begin{figure}[htbp]
	\centering
	\includegraphics[width=0.45\textwidth]{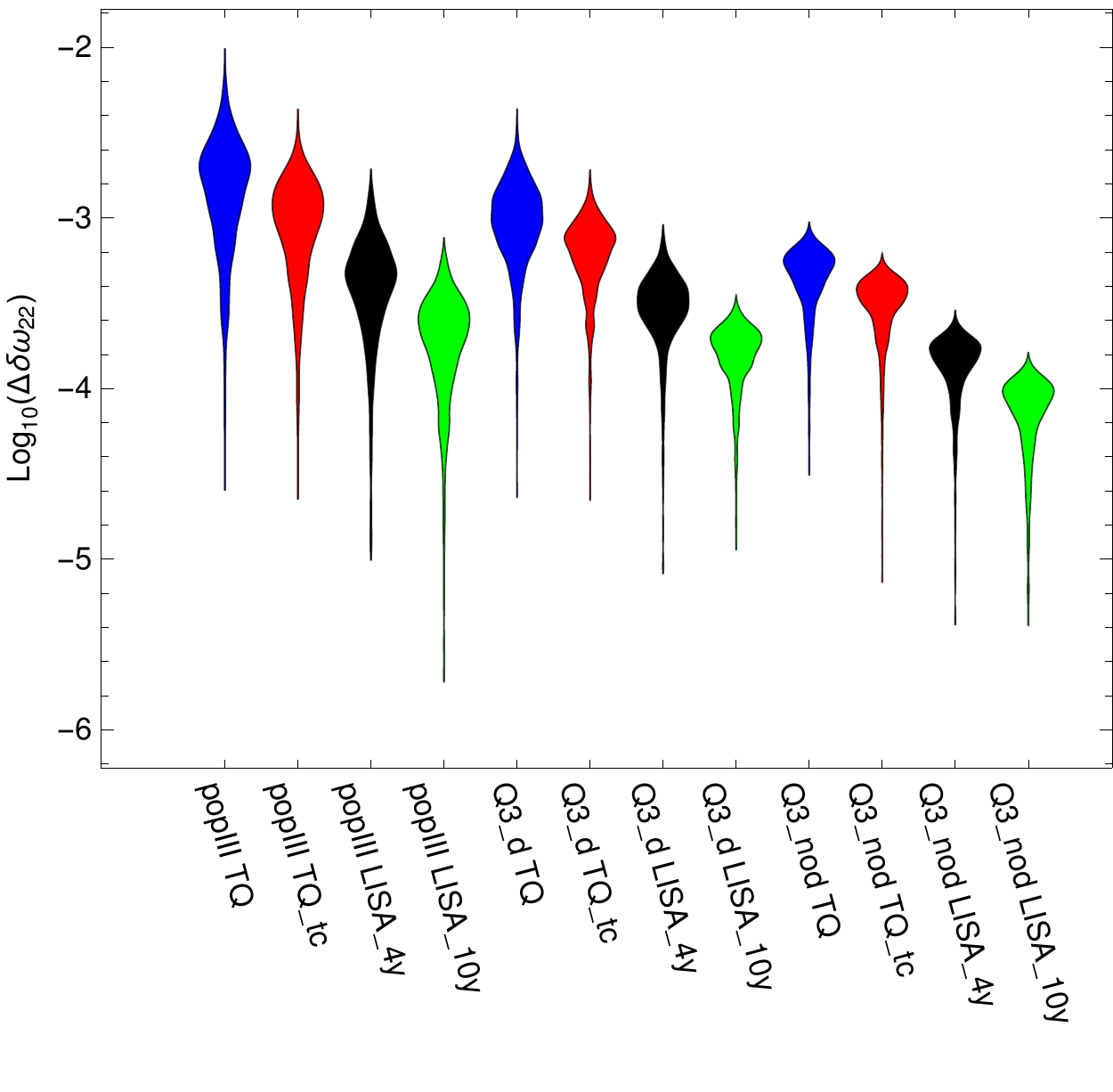}
	\caption{Violin plots showing the distribution of  uncertainty in the parameter estimation of $\delta\omega_{22}$ under different detector scenarios and different astrophysical models, as labeled on the x-axis.}\label{figboxw22}
\end{figure}

\begin{figure}[htbp]
	\centering
	\includegraphics[width=0.45\textwidth]{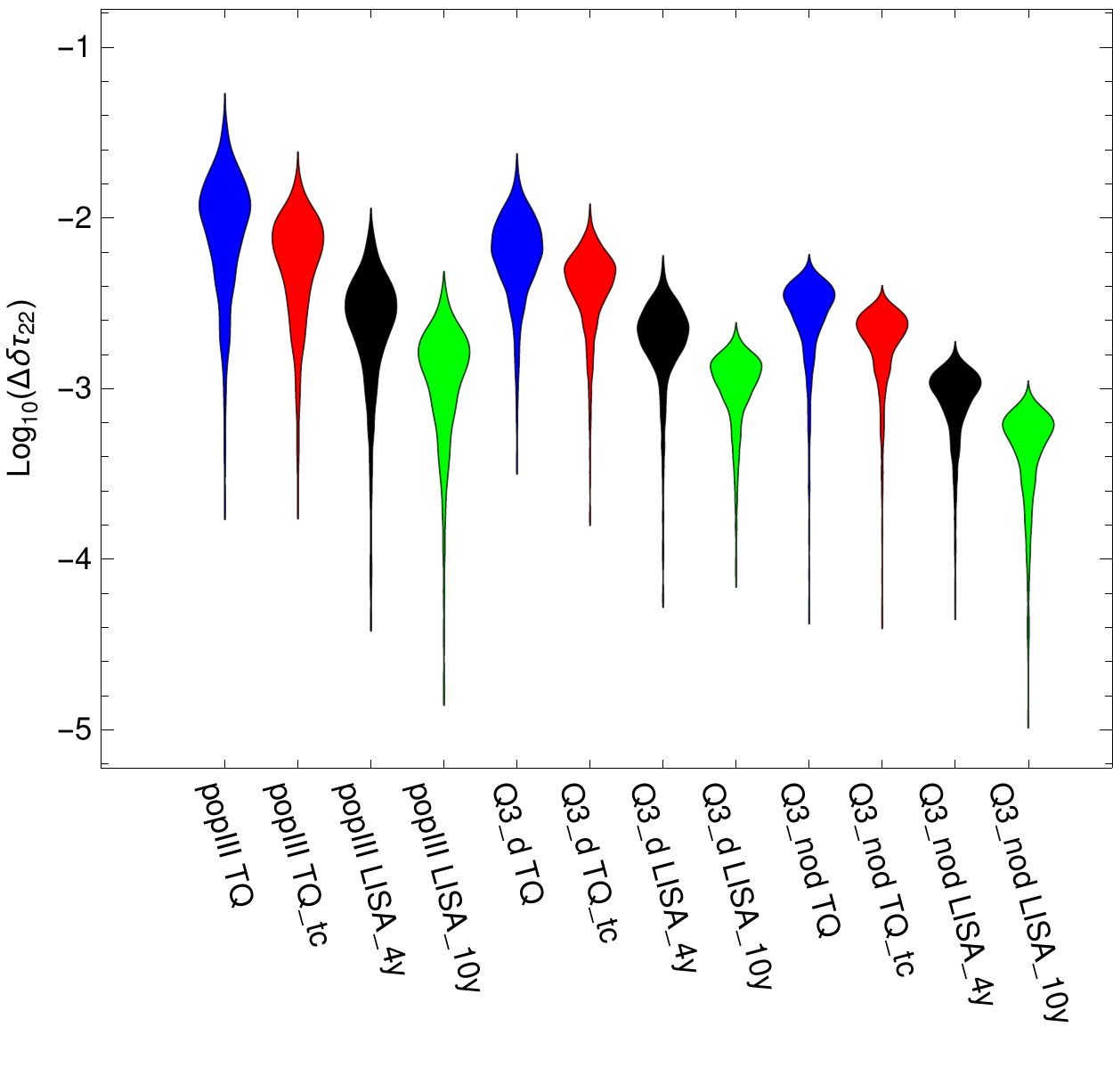}
	\caption{Same as figure \ref{figboxw22} but for $\delta\tau_{22}$.}\label{figboxt22}
\end{figure}

\begin{figure}[htbp]
	\centering
	\includegraphics[width=0.45\textwidth]{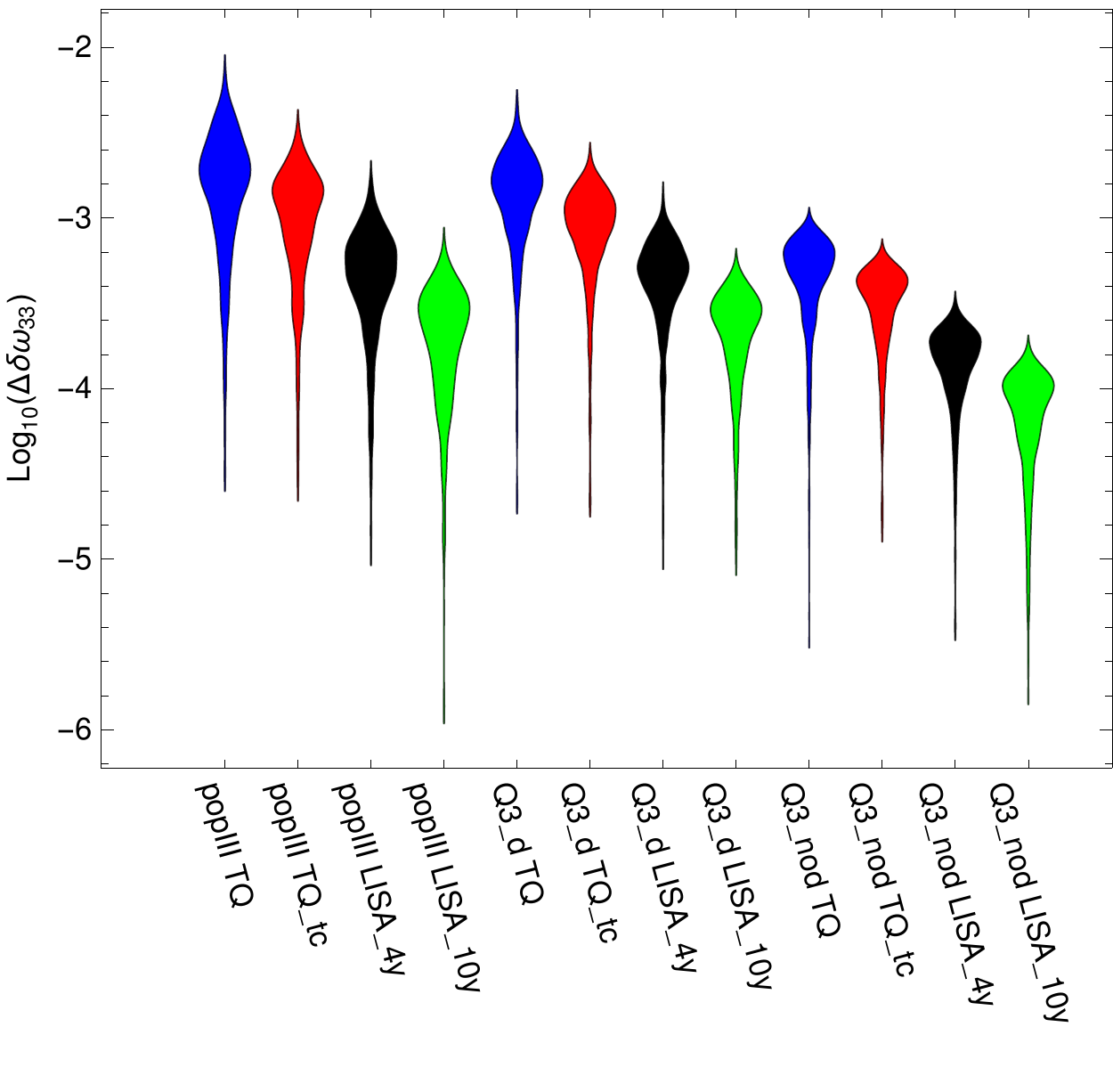}
	\caption{Same as figure \ref{figboxw22} but for $\delta\omega_{33}$.}\label{figboxw33}
\end{figure}

The results on expected constraints on the no-hair theorem violating parameters \(\delta\omega_{22}\), \(\delta\tau_{22}\) and \(\delta\omega_{33}\) are presented in figure \ref{figboxw22}, \ref{figboxt22} and \ref{figboxw33} respectively, and summarized in Table \ref{tablemr}. Once fixed the  detector and the MBH scenario, each of the 1000 mock catalogs will result in a different number of testing events with different properties, and thus in a different constraint on the parameters. The distribution of these combined constraints over the 1000 mock catalogs is what is shown by means of 'violin plots' in the three figures. Conversely, the table reports the mean value of the constraints over the 1000 realizations, together with the standard deviation.

We first notice that GW detectors are more sensitive to deviations in characteristic frequencies (\(\delta\omega_{22}\),  and \(\delta\omega_{33}\)) rather than damping time \(\delta\tau_{22}\). The former being generally constrained about five time better than the latter. Constraints also strongly depend on the assumed MBH population model. Despite resulting in a comparable number of testing events, the Q3\_d model provides constraints that are 2-to-3 times tighter than the popIII one. This is because GW sources are generally more massive in the Q3\_d model, and the ringdown is detected for almost all the events at high SNR, which is not the case for the popIII model. The even stronger constraints provided by the Q3\_nod models are due to the larger number of testing events. Note, however, that compared to the Q3\_nod model the improvement does not scale with $\sqrt{N_{\rm Testing}}$, as one would naively expect. This is because in the Q3\_d model, the MBH binary dynamics is taken into account, resulting in long merger timescales that push the distribution of observed systems at lower redshift compared to the Q3\_nod model. Therefore, there are less events available for testing purposes, but they have larger SNR on average. Finally, LISA generally provides a factor 3-to-5 better constraints then TQ on all parameters. Again, this is partly due to the larger number of detected systems, but also by the fact that sources detected by LISA have typically higher SNR than those seen by TQ (cf figure \ref{figmass}). Focusing on TQ, depending on the detector configuration and MBH population model, the decay parameter \(\delta\tau_{22}\) can be constrained to 0.2\% to 1.5\% accuracy; these numbers become a factor of $\approx10$ better for the frequency parameters (\(\delta\omega_{22}\),  and \(\delta\omega_{33}\)) that can both be constrained within 0.03\% to 0.3\%.

\begin{table*}[!htbp]
	\label{tablemr}
	\begin{tabular}{|p{2.5cm}|p{1.6cm}<{\centering}|p{1.6cm}<{\centering}|p{2cm}<{\centering}|p{3cm}<{\centering}|p{3cm}<{\centering}|p{3cm}<{\centering}|}
		\hline
		\cline{1-6}
		Cases&$N_{\rho_{\rm IMR}>8}$&$N_{\rho_{\rm Rd}>8}$&$N_{\rho_{\rm Rd}>\rho_{\rm GLRT}}$& \(\Delta\delta\omega_{22}\)& \(\Delta\delta\tau_{22}\)& \(\Delta\delta\omega_{33}\)\\
		\hline
		pop\MyRoman{3} TQ&51.7&16.5&12.7&\(0.0023\pm0.0014\)&\(0.015\pm0.0092\)&\(0.0029\pm0.0019\)\\
		\hline
		pop\MyRoman{3} TQ\_tc&104.0&31.4&24.3&\(0.0013\pm0.00079\)&\(0.0085\pm0.0051\)&\(0.0016\pm0.0011\)\\
		\hline
		pop\MyRoman{3} LISA\_4y&118.5&28.9&23.0&\(0.00046\pm0.00027\)&\(0.0028\pm0.0017\)&\(0.00051\pm0.00031\)\\
		\hline
        pop\MyRoman{3} LISA\_10y&296.6&72.9&58.5&\(0.00021\pm0.00012\)&\(0.0013\pm0.00076\)&\(0.00023\pm0.00014\)\\
        \hline		
		Q3\_d TQ&17.7&17.2&15.1&\(0.00080\pm0.00041\)&\(0.0052\pm0.0027\)&\(0.0014\pm0.00096\)\\
		\hline
		Q3\_d TQ\_tc&35.7&34.3&29.4&\(0.00052\pm0.00023\)&\(0.0034\pm0.0015\)&\(0.00078\pm0.00043\)\\
		\hline
		Q3\_d LISA\_4y&29.7&28.8&24.3&\(0.00032\pm0.00014\)&\(0.0021\pm0.00091\)&\(0.00047\pm0.00024\)\\
		\hline
		Q3\_d LISA\_10y&75.5&73.6&62.0&\(0.00016\pm0.000061\)&\(0.0011\pm0.00043\)&\(0.00023\pm0.00011\)\\
		\hline
		Q3\_nod TQ&274.9&247.5&162.1&\(0.00044\pm0.00017\)&\(0.0027\pm0.0011\)&\(0.00041\pm0.00021\)\\
		\hline
		Q3\_nod TQ\_tc&535.4&486.2&317.8&\(0.00033\pm0.00011\)&\(0.0021\pm0.00073\)&\(0.00031\pm0.00014\)\\
		\hline	
		Q3\_nod LISA\_4y&441.8&399.7&261.4&\(0.00014\pm0.000051\)&\(0.00089\pm0.00034\)&\(0.00015\pm0.000065\)\\
		\hline
		Q3\_nod LISA\_10y&1102.8&997.8&652.1&\(0.000075\pm0.000031\)&\(0.00047\pm0.00021\)&\(0.000074\pm0.000036\)\\
		\hline
	\end{tabular}
\caption{This table shows the average IMR detection number ($N_{\rho_{\rm IMR}>8}$), the ringdown detection number ($N_{\rho_{\rm Rd}>8}$), the average testing number ($N_{\rho_{\rm Rd}>\rho_{\rm GLRT}}$) and the average constraint on \(\delta\omega_{22}\), \(\delta\tau_{22}\) and \(\delta\omega_{33}\) under different detector scenarios and different astrophysical models.}
\end{table*}

\section{Summary and future work}\label{summary}

In this paper, we have studied TianQin's capability to test the no-hair theorem.

We have modeled the waveform of the ringdown signal from the merger of a massive black hole binary, by including the four strongest QNMs of the remnant Kerr black hole. We then used a set of phenomenological parameters modifying the frequencies and the damping times of the QNMs to parametrize the effect of a no-hair theorem violation. We further assumed that the no-hair theorem violating parameters are independent of the other source parameters.

We have used the Fisher information matrix method to study how these parameters are constrained by a single detection of a massive black hole merger by TianQin. We have studied how the constraints on the no-hair theorem violating parameters vary with the observed mass, luminosity distance, final spin, symmetric mass ratio and effective spin of the source.

We have found that \(\delta\omega_{22}\), \(\delta\tau_{22}\) and \(\delta\omega_{33}\) are the best constrained parameters in the majority of cases. For a single detection, we find that TianQin and LISA  provide  constraints on those three parameters in different mass ranges: although LISA can extract more information from binaries with $M_z>3\times 10^6$M$_\odot$, TianQin is better suited to test no-hair theorem using binaries with $M_z<10^6$M$_\odot$.
Joint detections with TianQin and LISA will further improve no-hair theorem tests for binaries of about $10^6$M$_\odot$, where the performance of the two detectors is comparable. By combining constraints from all the events expected throughout the lifetime of TianQin,  \(\delta\omega_{22}\), \(\delta\tau_{22}\) and \(\delta\omega_{33}\) can be constrained  to within $0.0004\sim0.002$, $0.002\sim0.01$ and $0.0004\sim0.003$ respectively, depending on the massive black hole population model (see Table \ref{tablemr}).

For completeness, we have also considered other detector scenarios, including a twin set of TianQin constellations running for 5 years, and LISA running for 4 years or 10 years.
Running a twin set of TianQin trivially improves the constrains by about $\sqrt{2}$. LISA, on the other hand, can detect more massive binaries at higher SNR, thus offering the possibility of detecting smaller deviations from the no-hair theorem. Constraints on \(\delta\tau_{22}\) range between $0.003$ and $0.0005$ depending on the massive black hole population model and on the duration of the LISA mission (cf. Table \ref{tablemr}).

We note that the present work can be improved in many directions, for example by including the effect of eccentricity of the progenitor binary, by considering more than four QNMs and other GW polarizations in the ringdown signal, and by using more robust parameter estimation methods than the Fisher information matrix. One should also consider how to relate the phenomenological parameters used in this paper to the parameters of a specific theory of gravity which predicting violations of the no-hair theorem.
We will provide an explicit example of this latter point in \cite{Bao:2019kgt}, for the
special case of
 Scalar-Tensor-Vector Gravity.

\begin{acknowledgments}
The authors thank Shun-Jia Huang, Peng-Cheng Li, Xian-Ji Ye,Yi-Fan Wang and John Veitch for useful discussion. We also thank to Gregorio Carullo for kind comments and suggestions.  This work has been supported by the Natural Science Foundation of China (Grant Nos. 11805286, 11703098, 91636111,  11690022,  11475064) and by the European Union's Horizon 2020 research and innovation program under the Marie Sklodowska-Curie grant agreement No 690904. This project has received funding (to E. Barausse) from the European Research Council (ERC) under the European Union¡¯s Horizon 2020
research and innovation programme (grant agreement no. GRAMS-815673; project title ``GRavity from Astrophysical to Microscopic Scales''). AS is supported by the Royal Society.
\end{acknowledgments}

\bibliographystyle{apsrev4-1}

\bibliography{TQ-QNM}

\begin{thebibliography}{78}%
\makeatletter
\providecommand \@ifxundefined [1]{%
 \@ifx{#1\undefined}
}%
\providecommand \@ifnum [1]{%
 \ifnum #1\expandafter \@firstoftwo
 \else \expandafter \@secondoftwo
 \fi
}%
\providecommand \@ifx [1]{%
 \ifx #1\expandafter \@firstoftwo
 \else \expandafter \@secondoftwo
 \fi
}%
\providecommand \natexlab [1]{#1}%
\providecommand \enquote  [1]{``#1''}%
\providecommand \bibnamefont  [1]{#1}%
\providecommand \bibfnamefont [1]{#1}%
\providecommand \citenamefont [1]{#1}%
\providecommand \href@noop [0]{\@secondoftwo}%
\providecommand \href [0]{\begingroup \@sanitize@url \@href}%
\providecommand \@href[1]{\@@startlink{#1}\@@href}%
\providecommand \@@href[1]{\endgroup#1\@@endlink}%
\providecommand \@sanitize@url [0]{\catcode `\\12\catcode `\$12\catcode
  `\&12\catcode `\#12\catcode `\^12\catcode `\_12\catcode `\%12\relax}%
\providecommand \@@startlink[1]{}%
\providecommand \@@endlink[0]{}%
\providecommand \url  [0]{\begingroup\@sanitize@url \@url }%
\providecommand \@url [1]{\endgroup\@href {#1}{\urlprefix }}%
\providecommand \urlprefix  [0]{URL }%
\providecommand \Eprint [0]{\href }%
\providecommand \doibase [0]{http://dx.doi.org/}%
\providecommand \selectlanguage [0]{\@gobble}%
\providecommand \bibinfo  [0]{\@secondoftwo}%
\providecommand \bibfield  [0]{\@secondoftwo}%
\providecommand \translation [1]{[#1]}%
\providecommand \BibitemOpen [0]{}%
\providecommand \bibitemStop [0]{}%
\providecommand \bibitemNoStop [0]{.\EOS\space}%
\providecommand \EOS [0]{\spacefactor3000\relax}%
\providecommand \BibitemShut  [1]{\csname bibitem#1\endcsname}%
\let\auto@bib@innerbib\@empty
\bibitem [{\citenamefont {Abbott}\ \emph
  {et~al.}(2016{\natexlab{a}})\citenamefont {Abbott} \emph
  {et~al.}}]{Abbott:2016blz}%
  \BibitemOpen
  \bibfield  {author} {\bibinfo {author} {\bibfnamefont {B.~P.}\ \bibnamefont
  {Abbott}} \emph {et~al.} (\bibinfo {collaboration} {Virgo, LIGO
  Scientific}),\ }\href {\doibase 10.1103/PhysRevLett.116.061102} {\bibfield
  {journal} {\bibinfo  {journal} {Phys. Rev. Lett.}\ }\textbf {\bibinfo
  {volume} {116}},\ \bibinfo {pages} {061102} (\bibinfo {year}
  {2016}{\natexlab{a}})},\ \Eprint {http://arxiv.org/abs/1602.03837}
  {arXiv:1602.03837 [gr-qc]} \BibitemShut {NoStop}%
\bibitem [{\citenamefont {Abbott}\ \emph
  {et~al.}(2016{\natexlab{b}})\citenamefont {Abbott} \emph
  {et~al.}}]{Abbott:2016nmj}%
  \BibitemOpen
  \bibfield  {author} {\bibinfo {author} {\bibfnamefont {B.~P.}\ \bibnamefont
  {Abbott}} \emph {et~al.} (\bibinfo {collaboration} {Virgo, LIGO
  Scientific}),\ }\href {\doibase 10.1103/PhysRevLett.116.241103} {\bibfield
  {journal} {\bibinfo  {journal} {Phys. Rev. Lett.}\ }\textbf {\bibinfo
  {volume} {116}},\ \bibinfo {pages} {241103} (\bibinfo {year}
  {2016}{\natexlab{b}})},\ \Eprint {http://arxiv.org/abs/1606.04855}
  {arXiv:1606.04855 [gr-qc]} \BibitemShut {NoStop}%
\bibitem [{\citenamefont {Abbott}\ \emph
  {et~al.}(2017{\natexlab{a}})\citenamefont {Abbott} \emph
  {et~al.}}]{Abbott:2017gyy}%
  \BibitemOpen
  \bibfield  {author} {\bibinfo {author} {\bibfnamefont {B.~P.}\ \bibnamefont
  {Abbott}} \emph {et~al.} (\bibinfo {collaboration} {Virgo, LIGO
  Scientific}),\ }\href {\doibase 10.3847/2041-8213/aa9f0c} {\bibfield
  {journal} {\bibinfo  {journal} {Astrophys. J.}\ }\textbf {\bibinfo {volume}
  {851}},\ \bibinfo {pages} {L35} (\bibinfo {year} {2017}{\natexlab{a}})},\
  \Eprint {http://arxiv.org/abs/1711.05578} {arXiv:1711.05578 [astro-ph.HE]}
  \BibitemShut {NoStop}%
\bibitem [{\citenamefont {Abbott}\ \emph
  {et~al.}(2017{\natexlab{b}})\citenamefont {Abbott} \emph
  {et~al.}}]{Abbott:2017oio}%
  \BibitemOpen
  \bibfield  {author} {\bibinfo {author} {\bibfnamefont {B.~P.}\ \bibnamefont
  {Abbott}} \emph {et~al.} (\bibinfo {collaboration} {Virgo, LIGO
  Scientific}),\ }\href {\doibase 10.1103/PhysRevLett.119.141101} {\bibfield
  {journal} {\bibinfo  {journal} {Phys. Rev. Lett.}\ }\textbf {\bibinfo
  {volume} {119}},\ \bibinfo {pages} {141101} (\bibinfo {year}
  {2017}{\natexlab{b}})},\ \Eprint {http://arxiv.org/abs/1709.09660}
  {arXiv:1709.09660 [gr-qc]} \BibitemShut {NoStop}%
\bibitem [{\citenamefont {Abbott}\ \emph
  {et~al.}(2017{\natexlab{c}})\citenamefont {Abbott} \emph
  {et~al.}}]{Abbott:2017vtc}%
  \BibitemOpen
  \bibfield  {author} {\bibinfo {author} {\bibfnamefont {B.~P.}\ \bibnamefont
  {Abbott}} \emph {et~al.} (\bibinfo {collaboration} {VIRGO, LIGO
  Scientific}),\ }\href {\doibase 10.1103/PhysRevLett.118.221101,
  10.1103/PhysRevLett.121.129901} {\bibfield  {journal} {\bibinfo  {journal}
  {Phys. Rev. Lett.}\ }\textbf {\bibinfo {volume} {118}},\ \bibinfo {pages}
  {221101} (\bibinfo {year} {2017}{\natexlab{c}})},\ \bibinfo {note} {[Erratum:
  Phys. Rev. Lett.121,no.12,129901(2018)]},\ \Eprint
  {http://arxiv.org/abs/1706.01812} {arXiv:1706.01812 [gr-qc]} \BibitemShut
  {NoStop}%
\bibitem [{\citenamefont {Abbott}\ \emph
  {et~al.}(2016{\natexlab{c}})\citenamefont {Abbott} \emph
  {et~al.}}]{TheLIGOScientific:2016pea}%
  \BibitemOpen
  \bibfield  {author} {\bibinfo {author} {\bibfnamefont {B.~P.}\ \bibnamefont
  {Abbott}} \emph {et~al.} (\bibinfo {collaboration} {Virgo, LIGO
  Scientific}),\ }\href {\doibase 10.1103/PhysRevX.6.041015,
  10.1103/PhysRevX.8.039903} {\bibfield  {journal} {\bibinfo  {journal} {Phys.
  Rev.}\ }\textbf {\bibinfo {volume} {X6}},\ \bibinfo {pages} {041015}
  (\bibinfo {year} {2016}{\natexlab{c}})},\ \bibinfo {note} {[Erratum: Phys.
  Rev.X8,no.3,039903(2018)]},\ \Eprint {http://arxiv.org/abs/1606.04856}
  {arXiv:1606.04856 [gr-qc]} \BibitemShut {NoStop}%
\bibitem [{\citenamefont {Abbott}\ \emph
  {et~al.}(2017{\natexlab{d}})\citenamefont {Abbott} \emph
  {et~al.}}]{TheLIGOScientific:2017qsa}%
  \BibitemOpen
  \bibfield  {author} {\bibinfo {author} {\bibfnamefont {B.}~\bibnamefont
  {Abbott}} \emph {et~al.} (\bibinfo {collaboration} {Virgo, LIGO
  Scientific}),\ }\href {\doibase 10.1103/PhysRevLett.119.161101} {\bibfield
  {journal} {\bibinfo  {journal} {Phys. Rev. Lett.}\ }\textbf {\bibinfo
  {volume} {119}},\ \bibinfo {pages} {161101} (\bibinfo {year}
  {2017}{\natexlab{d}})},\ \Eprint {http://arxiv.org/abs/1710.05832}
  {arXiv:1710.05832 [gr-qc]} \BibitemShut {NoStop}%
\bibitem [{\citenamefont {Abbott}\ \emph
  {et~al.}(2016{\natexlab{d}})\citenamefont {Abbott} \emph
  {et~al.}}]{TheLIGOScientific:2016src}%
  \BibitemOpen
  \bibfield  {author} {\bibinfo {author} {\bibfnamefont {B.~P.}\ \bibnamefont
  {Abbott}} \emph {et~al.} (\bibinfo {collaboration} {Virgo, LIGO
  Scientific}),\ }\href {\doibase 10.1103/PhysRevLett.116.221101,
  10.1103/PhysRevLett.121.129902} {\bibfield  {journal} {\bibinfo  {journal}
  {Phys. Rev. Lett.}\ }\textbf {\bibinfo {volume} {116}},\ \bibinfo {pages}
  {221101} (\bibinfo {year} {2016}{\natexlab{d}})},\ \bibinfo {note} {[Erratum:
  Phys. Rev. Lett.121,no.12,129902(2018)]},\ \Eprint
  {http://arxiv.org/abs/1602.03841} {arXiv:1602.03841 [gr-qc]} \BibitemShut
  {NoStop}%
\bibitem [{\citenamefont {Berti}\ \emph
  {et~al.}(2018{\natexlab{a}})\citenamefont {Berti}, \citenamefont {Yagi},\
  and\ \citenamefont {Yunes}}]{Berti:2018cxi}%
  \BibitemOpen
  \bibfield  {author} {\bibinfo {author} {\bibfnamefont {E.}~\bibnamefont
  {Berti}}, \bibinfo {author} {\bibfnamefont {K.}~\bibnamefont {Yagi}}, \ and\
  \bibinfo {author} {\bibfnamefont {N.}~\bibnamefont {Yunes}},\ }\href
  {\doibase 10.1007/s10714-018-2362-8} {\bibfield  {journal} {\bibinfo
  {journal} {Gen. Rel. Grav.}\ }\textbf {\bibinfo {volume} {50}},\ \bibinfo
  {pages} {46} (\bibinfo {year} {2018}{\natexlab{a}})},\ \Eprint
  {http://arxiv.org/abs/1801.03208} {arXiv:1801.03208 [gr-qc]} \BibitemShut
  {NoStop}%
\bibitem [{\citenamefont {Berti}\ \emph
  {et~al.}(2018{\natexlab{b}})\citenamefont {Berti}, \citenamefont {Yagi},
  \citenamefont {Yang},\ and\ \citenamefont {Yunes}}]{Berti:2018vdi}%
  \BibitemOpen
  \bibfield  {author} {\bibinfo {author} {\bibfnamefont {E.}~\bibnamefont
  {Berti}}, \bibinfo {author} {\bibfnamefont {K.}~\bibnamefont {Yagi}},
  \bibinfo {author} {\bibfnamefont {H.}~\bibnamefont {Yang}}, \ and\ \bibinfo
  {author} {\bibfnamefont {N.}~\bibnamefont {Yunes}},\ }\href {\doibase
  10.1007/s10714-018-2372-6} {\bibfield  {journal} {\bibinfo  {journal} {Gen.
  Rel. Grav.}\ }\textbf {\bibinfo {volume} {50}},\ \bibinfo {pages} {49}
  (\bibinfo {year} {2018}{\natexlab{b}})},\ \Eprint
  {http://arxiv.org/abs/1801.03587} {arXiv:1801.03587 [gr-qc]} \BibitemShut
  {NoStop}%
\bibitem [{\citenamefont {Alexander}\ and\ \citenamefont
  {Yunes}(2018)}]{Alexander:2017jmt}%
  \BibitemOpen
  \bibfield  {author} {\bibinfo {author} {\bibfnamefont {S.~H.}\ \bibnamefont
  {Alexander}}\ and\ \bibinfo {author} {\bibfnamefont {N.}~\bibnamefont
  {Yunes}},\ }\href {\doibase 10.1103/PhysRevD.97.064033} {\bibfield  {journal}
  {\bibinfo  {journal} {Phys. Rev.}\ }\textbf {\bibinfo {volume} {D97}},\
  \bibinfo {pages} {064033} (\bibinfo {year} {2018})},\ \Eprint
  {http://arxiv.org/abs/1712.01853} {arXiv:1712.01853 [gr-qc]} \BibitemShut
  {NoStop}%
\bibitem [{\citenamefont {Chamberlain}\ and\ \citenamefont
  {Yunes}(2017)}]{Chamberlain:2017fjl}%
  \BibitemOpen
  \bibfield  {author} {\bibinfo {author} {\bibfnamefont {K.}~\bibnamefont
  {Chamberlain}}\ and\ \bibinfo {author} {\bibfnamefont {N.}~\bibnamefont
  {Yunes}},\ }\href {\doibase 10.1103/PhysRevD.96.084039} {\bibfield  {journal}
  {\bibinfo  {journal} {Phys. Rev.}\ }\textbf {\bibinfo {volume} {D96}},\
  \bibinfo {pages} {084039} (\bibinfo {year} {2017})},\ \Eprint
  {http://arxiv.org/abs/1704.08268} {arXiv:1704.08268 [gr-qc]} \BibitemShut
  {NoStop}%
\bibitem [{\citenamefont {Yunes}\ \emph {et~al.}(2016)\citenamefont {Yunes},
  \citenamefont {Yagi},\ and\ \citenamefont {Pretorius}}]{Yunes:2016jcc}%
  \BibitemOpen
  \bibfield  {author} {\bibinfo {author} {\bibfnamefont {N.}~\bibnamefont
  {Yunes}}, \bibinfo {author} {\bibfnamefont {K.}~\bibnamefont {Yagi}}, \ and\
  \bibinfo {author} {\bibfnamefont {F.}~\bibnamefont {Pretorius}},\ }\href
  {\doibase 10.1103/PhysRevD.94.084002} {\bibfield  {journal} {\bibinfo
  {journal} {Phys. Rev.}\ }\textbf {\bibinfo {volume} {D94}},\ \bibinfo {pages}
  {084002} (\bibinfo {year} {2016})},\ \Eprint
  {http://arxiv.org/abs/1603.08955} {arXiv:1603.08955 [gr-qc]} \BibitemShut
  {NoStop}%
\bibitem [{\citenamefont {Barausse}\ \emph {et~al.}(2016)\citenamefont
  {Barausse}, \citenamefont {Yunes},\ and\ \citenamefont
  {Chamberlain}}]{Barausse:2016eii}%
  \BibitemOpen
  \bibfield  {author} {\bibinfo {author} {\bibfnamefont {E.}~\bibnamefont
  {Barausse}}, \bibinfo {author} {\bibfnamefont {N.}~\bibnamefont {Yunes}}, \
  and\ \bibinfo {author} {\bibfnamefont {K.}~\bibnamefont {Chamberlain}},\
  }\href {\doibase 10.1103/PhysRevLett.116.241104} {\bibfield  {journal}
  {\bibinfo  {journal} {Phys. Rev. Lett.}\ }\textbf {\bibinfo {volume} {116}},\
  \bibinfo {pages} {241104} (\bibinfo {year} {2016})},\ \Eprint
  {http://arxiv.org/abs/1603.04075} {arXiv:1603.04075 [gr-qc]} \BibitemShut
  {NoStop}%
\bibitem [{\citenamefont {Chatziioannou}\ \emph {et~al.}(2015)\citenamefont
  {Chatziioannou}, \citenamefont {Yagi}, \citenamefont {Klein}, \citenamefont
  {Cornish},\ and\ \citenamefont {Yunes}}]{Chatziioannou:2015uea}%
  \BibitemOpen
  \bibfield  {author} {\bibinfo {author} {\bibfnamefont {K.}~\bibnamefont
  {Chatziioannou}}, \bibinfo {author} {\bibfnamefont {K.}~\bibnamefont {Yagi}},
  \bibinfo {author} {\bibfnamefont {A.}~\bibnamefont {Klein}}, \bibinfo
  {author} {\bibfnamefont {N.}~\bibnamefont {Cornish}}, \ and\ \bibinfo
  {author} {\bibfnamefont {N.}~\bibnamefont {Yunes}},\ }\href {\doibase
  10.1103/PhysRevD.92.104008} {\bibfield  {journal} {\bibinfo  {journal} {Phys.
  Rev.}\ }\textbf {\bibinfo {volume} {D92}},\ \bibinfo {pages} {104008}
  (\bibinfo {year} {2015})},\ \Eprint {http://arxiv.org/abs/1508.02062}
  {arXiv:1508.02062 [gr-qc]} \BibitemShut {NoStop}%
\bibitem [{\citenamefont {Israel}(1967)}]{Israel:1967wq}%
  \BibitemOpen
  \bibfield  {author} {\bibinfo {author} {\bibfnamefont {W.}~\bibnamefont
  {Israel}},\ }\href {\doibase 10.1103/PhysRev.164.1776} {\bibfield  {journal}
  {\bibinfo  {journal} {Phys. Rev.}\ }\textbf {\bibinfo {volume} {164}},\
  \bibinfo {pages} {1776} (\bibinfo {year} {1967})}\BibitemShut {NoStop}%
\bibitem [{\citenamefont {Israel}(1968)}]{Israel:1967za}%
  \BibitemOpen
  \bibfield  {author} {\bibinfo {author} {\bibfnamefont {W.}~\bibnamefont
  {Israel}},\ }\href {\doibase 10.1007/BF01645859} {\bibfield  {journal}
  {\bibinfo  {journal} {Commun. Math. Phys.}\ }\textbf {\bibinfo {volume}
  {8}},\ \bibinfo {pages} {245} (\bibinfo {year} {1968})}\BibitemShut {NoStop}%
\bibitem [{\citenamefont {Carter}(1971)}]{Carter:1971zc}%
  \BibitemOpen
  \bibfield  {author} {\bibinfo {author} {\bibfnamefont {B.}~\bibnamefont
  {Carter}},\ }\href {\doibase 10.1103/PhysRevLett.26.331} {\bibfield
  {journal} {\bibinfo  {journal} {Phys. Rev. Lett.}\ }\textbf {\bibinfo
  {volume} {26}},\ \bibinfo {pages} {331} (\bibinfo {year} {1971})}\BibitemShut
  {NoStop}%
\bibitem [{\citenamefont {Gibbons}(1975)}]{Gibbons:1975kk}%
  \BibitemOpen
  \bibfield  {author} {\bibinfo {author} {\bibfnamefont {G.~W.}\ \bibnamefont
  {Gibbons}},\ }\href {\doibase 10.1007/BF01609829} {\bibfield  {journal}
  {\bibinfo  {journal} {Commun. Math. Phys.}\ }\textbf {\bibinfo {volume}
  {44}},\ \bibinfo {pages} {245} (\bibinfo {year} {1975})}\BibitemShut
  {NoStop}%
\bibitem [{\citenamefont {Hanni}(1982)}]{hanni1982limits}%
  \BibitemOpen
  \bibfield  {author} {\bibinfo {author} {\bibfnamefont {R.}~\bibnamefont
  {Hanni}},\ }\href@noop {} {\bibfield  {journal} {\bibinfo  {journal}
  {Physical Review D}\ }\textbf {\bibinfo {volume} {25}},\ \bibinfo {pages}
  {2509} (\bibinfo {year} {1982})}\BibitemShut {NoStop}%
\bibitem [{\citenamefont {Goldreich}\ and\ \citenamefont
  {Julian}(1969)}]{Goldreich:1969sb}%
  \BibitemOpen
  \bibfield  {author} {\bibinfo {author} {\bibfnamefont {P.}~\bibnamefont
  {Goldreich}}\ and\ \bibinfo {author} {\bibfnamefont {W.~H.}\ \bibnamefont
  {Julian}},\ }\href {\doibase 10.1086/150119} {\bibfield  {journal} {\bibinfo
  {journal} {Astrophys. J.}\ }\textbf {\bibinfo {volume} {157}},\ \bibinfo
  {pages} {869} (\bibinfo {year} {1969})}\BibitemShut {NoStop}%
\bibitem [{\citenamefont {Ruderman}\ and\ \citenamefont
  {Sutherland}(1975)}]{Ruderman:1975ju}%
  \BibitemOpen
  \bibfield  {author} {\bibinfo {author} {\bibfnamefont {M.~A.}\ \bibnamefont
  {Ruderman}}\ and\ \bibinfo {author} {\bibfnamefont {P.~G.}\ \bibnamefont
  {Sutherland}},\ }\href {\doibase 10.1086/153393} {\bibfield  {journal}
  {\bibinfo  {journal} {Astrophys. J.}\ }\textbf {\bibinfo {volume} {196}},\
  \bibinfo {pages} {51} (\bibinfo {year} {1975})}\BibitemShut {NoStop}%
\bibitem [{\citenamefont {Blandford}\ and\ \citenamefont
  {Znajek}(1977)}]{Blandford:1977ds}%
  \BibitemOpen
  \bibfield  {author} {\bibinfo {author} {\bibfnamefont {R.~D.}\ \bibnamefont
  {Blandford}}\ and\ \bibinfo {author} {\bibfnamefont {R.~L.}\ \bibnamefont
  {Znajek}},\ }\href {\doibase 10.1093/mnras/179.3.433} {\bibfield  {journal}
  {\bibinfo  {journal} {Mon. Not. Roy. Astron. Soc.}\ }\textbf {\bibinfo
  {volume} {179}},\ \bibinfo {pages} {433} (\bibinfo {year}
  {1977})}\BibitemShut {NoStop}%
\bibitem [{\citenamefont {Barausse}\ \emph {et~al.}(2014)\citenamefont
  {Barausse}, \citenamefont {Cardoso},\ and\ \citenamefont
  {Pani}}]{Barausse:2014tra}%
  \BibitemOpen
  \bibfield  {author} {\bibinfo {author} {\bibfnamefont {E.}~\bibnamefont
  {Barausse}}, \bibinfo {author} {\bibfnamefont {V.}~\bibnamefont {Cardoso}}, \
  and\ \bibinfo {author} {\bibfnamefont {P.}~\bibnamefont {Pani}},\ }\href
  {\doibase 10.1103/PhysRevD.89.104059} {\bibfield  {journal} {\bibinfo
  {journal} {Phys. Rev.}\ }\textbf {\bibinfo {volume} {D89}},\ \bibinfo {pages}
  {104059} (\bibinfo {year} {2014})},\ \Eprint {http://arxiv.org/abs/1404.7149}
  {arXiv:1404.7149 [gr-qc]} \BibitemShut {NoStop}%
\bibitem [{\citenamefont {Dreyer}\ \emph {et~al.}(2004)\citenamefont {Dreyer},
  \citenamefont {Kelly}, \citenamefont {Krishnan}, \citenamefont {Finn},
  \citenamefont {Garrison},\ and\ \citenamefont
  {Lopez-Aleman}}]{Dreyer:2003bv}%
  \BibitemOpen
  \bibfield  {author} {\bibinfo {author} {\bibfnamefont {O.}~\bibnamefont
  {Dreyer}}, \bibinfo {author} {\bibfnamefont {B.~J.}\ \bibnamefont {Kelly}},
  \bibinfo {author} {\bibfnamefont {B.}~\bibnamefont {Krishnan}}, \bibinfo
  {author} {\bibfnamefont {L.~S.}\ \bibnamefont {Finn}}, \bibinfo {author}
  {\bibfnamefont {D.}~\bibnamefont {Garrison}}, \ and\ \bibinfo {author}
  {\bibfnamefont {R.}~\bibnamefont {Lopez-Aleman}},\ }\href {\doibase
  10.1088/0264-9381/21/4/003} {\bibfield  {journal} {\bibinfo  {journal}
  {Class. Quant. Grav.}\ }\textbf {\bibinfo {volume} {21}},\ \bibinfo {pages}
  {787} (\bibinfo {year} {2004})},\ \Eprint
  {http://arxiv.org/abs/gr-qc/0309007} {arXiv:gr-qc/0309007 [gr-qc]}
  \BibitemShut {NoStop}%
\bibitem [{\citenamefont {Berti}\ \emph {et~al.}(2006)\citenamefont {Berti},
  \citenamefont {Cardoso},\ and\ \citenamefont {Will}}]{Berti:2005ys}%
  \BibitemOpen
  \bibfield  {author} {\bibinfo {author} {\bibfnamefont {E.}~\bibnamefont
  {Berti}}, \bibinfo {author} {\bibfnamefont {V.}~\bibnamefont {Cardoso}}, \
  and\ \bibinfo {author} {\bibfnamefont {C.~M.}\ \bibnamefont {Will}},\ }\href
  {\doibase 10.1103/PhysRevD.73.064030} {\bibfield  {journal} {\bibinfo
  {journal} {Phys. Rev.}\ }\textbf {\bibinfo {volume} {D73}},\ \bibinfo {pages}
  {064030} (\bibinfo {year} {2006})},\ \Eprint
  {http://arxiv.org/abs/gr-qc/0512160} {arXiv:gr-qc/0512160 [gr-qc]}
  \BibitemShut {NoStop}%
\bibitem [{\citenamefont {Berti}\ \emph {et~al.}(2007)\citenamefont {Berti},
  \citenamefont {Cardoso}, \citenamefont {Cardoso},\ and\ \citenamefont
  {Cavaglia}}]{Berti:2007zu}%
  \BibitemOpen
  \bibfield  {author} {\bibinfo {author} {\bibfnamefont {E.}~\bibnamefont
  {Berti}}, \bibinfo {author} {\bibfnamefont {J.}~\bibnamefont {Cardoso}},
  \bibinfo {author} {\bibfnamefont {V.}~\bibnamefont {Cardoso}}, \ and\
  \bibinfo {author} {\bibfnamefont {M.}~\bibnamefont {Cavaglia}},\ }\href
  {\doibase 10.1103/PhysRevD.76.104044} {\bibfield  {journal} {\bibinfo
  {journal} {Phys. Rev.}\ }\textbf {\bibinfo {volume} {D76}},\ \bibinfo {pages}
  {104044} (\bibinfo {year} {2007})},\ \Eprint {http://arxiv.org/abs/0707.1202}
  {arXiv:0707.1202 [gr-qc]} \BibitemShut {NoStop}%
\bibitem [{\citenamefont {Yang}\ \emph {et~al.}(2017)\citenamefont {Yang},
  \citenamefont {Yagi}, \citenamefont {Blackman}, \citenamefont {Lehner},
  \citenamefont {Paschalidis}, \citenamefont {Pretorius},\ and\ \citenamefont
  {Yunes}}]{Yang:2017zxs}%
  \BibitemOpen
  \bibfield  {author} {\bibinfo {author} {\bibfnamefont {H.}~\bibnamefont
  {Yang}}, \bibinfo {author} {\bibfnamefont {K.}~\bibnamefont {Yagi}}, \bibinfo
  {author} {\bibfnamefont {J.}~\bibnamefont {Blackman}}, \bibinfo {author}
  {\bibfnamefont {L.}~\bibnamefont {Lehner}}, \bibinfo {author} {\bibfnamefont
  {V.}~\bibnamefont {Paschalidis}}, \bibinfo {author} {\bibfnamefont
  {F.}~\bibnamefont {Pretorius}}, \ and\ \bibinfo {author} {\bibfnamefont
  {N.}~\bibnamefont {Yunes}},\ }\href {\doibase 10.1103/PhysRevLett.118.161101}
  {\bibfield  {journal} {\bibinfo  {journal} {Phys. Rev. Lett.}\ }\textbf
  {\bibinfo {volume} {118}},\ \bibinfo {pages} {161101} (\bibinfo {year}
  {2017})},\ \Eprint {http://arxiv.org/abs/1701.05808} {arXiv:1701.05808
  [gr-qc]} \BibitemShut {NoStop}%
\bibitem [{\citenamefont {Yang}\ \emph {et~al.}(2018)\citenamefont {Yang},
  \citenamefont {Paschalidis}, \citenamefont {Yagi}, \citenamefont {Lehner},
  \citenamefont {Pretorius},\ and\ \citenamefont {Yunes}}]{Yang:2017xlf}%
  \BibitemOpen
  \bibfield  {author} {\bibinfo {author} {\bibfnamefont {H.}~\bibnamefont
  {Yang}}, \bibinfo {author} {\bibfnamefont {V.}~\bibnamefont {Paschalidis}},
  \bibinfo {author} {\bibfnamefont {K.}~\bibnamefont {Yagi}}, \bibinfo {author}
  {\bibfnamefont {L.}~\bibnamefont {Lehner}}, \bibinfo {author} {\bibfnamefont
  {F.}~\bibnamefont {Pretorius}}, \ and\ \bibinfo {author} {\bibfnamefont
  {N.}~\bibnamefont {Yunes}},\ }\href {\doibase 10.1103/PhysRevD.97.024049}
  {\bibfield  {journal} {\bibinfo  {journal} {Phys. Rev.}\ }\textbf {\bibinfo
  {volume} {D97}},\ \bibinfo {pages} {024049} (\bibinfo {year} {2018})},\
  \Eprint {http://arxiv.org/abs/1707.00207} {arXiv:1707.00207 [gr-qc]}
  \BibitemShut {NoStop}%
\bibitem [{\citenamefont {Cunha}\ \emph {et~al.}(2017)\citenamefont {Cunha},
  \citenamefont {Berti},\ and\ \citenamefont {Herdeiro}}]{Cunha:2017qtt}%
  \BibitemOpen
  \bibfield  {author} {\bibinfo {author} {\bibfnamefont {P.~V.~P.}\
  \bibnamefont {Cunha}}, \bibinfo {author} {\bibfnamefont {E.}~\bibnamefont
  {Berti}}, \ and\ \bibinfo {author} {\bibfnamefont {C.~A.~R.}\ \bibnamefont
  {Herdeiro}},\ }\href {\doibase 10.1103/PhysRevLett.119.251102} {\bibfield
  {journal} {\bibinfo  {journal} {Phys. Rev. Lett.}\ }\textbf {\bibinfo
  {volume} {119}},\ \bibinfo {pages} {251102} (\bibinfo {year} {2017})},\
  \Eprint {http://arxiv.org/abs/1708.04211} {arXiv:1708.04211 [gr-qc]}
  \BibitemShut {NoStop}%
\bibitem [{\citenamefont {Barack}\ \emph {et~al.}(2018)\citenamefont {Barack}
  \emph {et~al.}}]{Barack:2018yly}%
  \BibitemOpen
  \bibfield  {author} {\bibinfo {author} {\bibfnamefont {L.}~\bibnamefont
  {Barack}} \emph {et~al.},\ }\href@noop {} {\  (\bibinfo {year} {2018})},\
  \Eprint {http://arxiv.org/abs/1806.05195} {arXiv:1806.05195 [gr-qc]}
  \BibitemShut {NoStop}%
\bibitem [{\citenamefont {Glampedakis}\ \emph {et~al.}(2017)\citenamefont
  {Glampedakis}, \citenamefont {Pappas}, \citenamefont {Silva},\ and\
  \citenamefont {Berti}}]{Glampedakis:2017dvb}%
  \BibitemOpen
  \bibfield  {author} {\bibinfo {author} {\bibfnamefont {K.}~\bibnamefont
  {Glampedakis}}, \bibinfo {author} {\bibfnamefont {G.}~\bibnamefont {Pappas}},
  \bibinfo {author} {\bibfnamefont {H.~O.}\ \bibnamefont {Silva}}, \ and\
  \bibinfo {author} {\bibfnamefont {E.}~\bibnamefont {Berti}},\ }\href
  {\doibase 10.1103/PhysRevD.96.064054} {\bibfield  {journal} {\bibinfo
  {journal} {Phys. Rev.}\ }\textbf {\bibinfo {volume} {D96}},\ \bibinfo {pages}
  {064054} (\bibinfo {year} {2017})},\ \Eprint
  {http://arxiv.org/abs/1706.07658} {arXiv:1706.07658 [gr-qc]} \BibitemShut
  {NoStop}%
\bibitem [{\citenamefont {Berti}\ \emph {et~al.}(2016)\citenamefont {Berti},
  \citenamefont {Sesana}, \citenamefont {Barausse}, \citenamefont {Cardoso},\
  and\ \citenamefont {Belczynski}}]{Berti:2016lat}%
  \BibitemOpen
  \bibfield  {author} {\bibinfo {author} {\bibfnamefont {E.}~\bibnamefont
  {Berti}}, \bibinfo {author} {\bibfnamefont {A.}~\bibnamefont {Sesana}},
  \bibinfo {author} {\bibfnamefont {E.}~\bibnamefont {Barausse}}, \bibinfo
  {author} {\bibfnamefont {V.}~\bibnamefont {Cardoso}}, \ and\ \bibinfo
  {author} {\bibfnamefont {K.}~\bibnamefont {Belczynski}},\ }\href {\doibase
  10.1103/PhysRevLett.117.101102} {\bibfield  {journal} {\bibinfo  {journal}
  {Phys. Rev. Lett.}\ }\textbf {\bibinfo {volume} {117}},\ \bibinfo {pages}
  {101102} (\bibinfo {year} {2016})},\ \Eprint
  {http://arxiv.org/abs/1605.09286} {arXiv:1605.09286 [gr-qc]} \BibitemShut
  {NoStop}%
\bibitem [{\citenamefont {Detweiler}(1980)}]{Detweiler:1980gk}%
  \BibitemOpen
  \bibfield  {author} {\bibinfo {author} {\bibfnamefont {S.~L.}\ \bibnamefont
  {Detweiler}},\ }\href {\doibase 10.1086/158109} {\bibfield  {journal}
  {\bibinfo  {journal} {Astrophys. J.}\ }\textbf {\bibinfo {volume} {239}},\
  \bibinfo {pages} {292} (\bibinfo {year} {1980})}\BibitemShut {NoStop}%
\bibitem [{\citenamefont {Kamaretsos}\ \emph
  {et~al.}(2012{\natexlab{a}})\citenamefont {Kamaretsos}, \citenamefont
  {Hannam}, \citenamefont {Husa},\ and\ \citenamefont
  {Sathyaprakash}}]{Kamaretsos:2011um}%
  \BibitemOpen
  \bibfield  {author} {\bibinfo {author} {\bibfnamefont {I.}~\bibnamefont
  {Kamaretsos}}, \bibinfo {author} {\bibfnamefont {M.}~\bibnamefont {Hannam}},
  \bibinfo {author} {\bibfnamefont {S.}~\bibnamefont {Husa}}, \ and\ \bibinfo
  {author} {\bibfnamefont {B.~S.}\ \bibnamefont {Sathyaprakash}},\ }\href
  {\doibase 10.1103/PhysRevD.85.024018} {\bibfield  {journal} {\bibinfo
  {journal} {Phys. Rev.}\ }\textbf {\bibinfo {volume} {D85}},\ \bibinfo {pages}
  {024018} (\bibinfo {year} {2012}{\natexlab{a}})},\ \Eprint
  {http://arxiv.org/abs/1107.0854} {arXiv:1107.0854 [gr-qc]} \BibitemShut
  {NoStop}%
\bibitem [{\citenamefont {Kamaretsos}\ \emph
  {et~al.}(2012{\natexlab{b}})\citenamefont {Kamaretsos}, \citenamefont
  {Hannam},\ and\ \citenamefont {Sathyaprakash}}]{Kamaretsos:2012bs}%
  \BibitemOpen
  \bibfield  {author} {\bibinfo {author} {\bibfnamefont {I.}~\bibnamefont
  {Kamaretsos}}, \bibinfo {author} {\bibfnamefont {M.}~\bibnamefont {Hannam}},
  \ and\ \bibinfo {author} {\bibfnamefont {B.}~\bibnamefont {Sathyaprakash}},\
  }\href {\doibase 10.1103/PhysRevLett.109.141102} {\bibfield  {journal}
  {\bibinfo  {journal} {Phys. Rev. Lett.}\ }\textbf {\bibinfo {volume} {109}},\
  \bibinfo {pages} {141102} (\bibinfo {year} {2012}{\natexlab{b}})},\ \Eprint
  {http://arxiv.org/abs/1207.0399} {arXiv:1207.0399 [gr-qc]} \BibitemShut
  {NoStop}%
\bibitem [{\citenamefont {Gossan}\ \emph {et~al.}(2012)\citenamefont {Gossan},
  \citenamefont {Veitch},\ and\ \citenamefont {Sathyaprakash}}]{Gossan:2011ha}%
  \BibitemOpen
  \bibfield  {author} {\bibinfo {author} {\bibfnamefont {S.}~\bibnamefont
  {Gossan}}, \bibinfo {author} {\bibfnamefont {J.}~\bibnamefont {Veitch}}, \
  and\ \bibinfo {author} {\bibfnamefont {B.~S.}\ \bibnamefont
  {Sathyaprakash}},\ }\href {\doibase 10.1103/PhysRevD.85.124056} {\bibfield
  {journal} {\bibinfo  {journal} {Phys. Rev.}\ }\textbf {\bibinfo {volume}
  {D85}},\ \bibinfo {pages} {124056} (\bibinfo {year} {2012})},\ \Eprint
  {http://arxiv.org/abs/1111.5819} {arXiv:1111.5819 [gr-qc]} \BibitemShut
  {NoStop}%
\bibitem [{\citenamefont {Amaro-Seoane}\ \emph {et~al.}(2012)\citenamefont
  {Amaro-Seoane} \emph {et~al.}}]{AmaroSeoane:2012je}%
  \BibitemOpen
  \bibfield  {author} {\bibinfo {author} {\bibfnamefont {P.}~\bibnamefont
  {Amaro-Seoane}} \emph {et~al.},\ }\href {\doibase
  10.1088/0264-9381/29/12/124016} {\bibfield  {journal} {\bibinfo  {journal}
  {Class. Quant. Grav.}\ }\textbf {\bibinfo {volume} {29}},\ \bibinfo {pages}
  {124016} (\bibinfo {year} {2012})},\ \Eprint {http://arxiv.org/abs/1202.0839}
  {arXiv:1202.0839 [gr-qc]} \BibitemShut {NoStop}%
\bibitem [{\citenamefont {Punturo}\ \emph {et~al.}(2010)\citenamefont {Punturo}
  \emph {et~al.}}]{Punturo:2010zz}%
  \BibitemOpen
  \bibfield  {author} {\bibinfo {author} {\bibfnamefont {M.}~\bibnamefont
  {Punturo}} \emph {et~al.},\ }\bibfield  {booktitle} {\emph {\bibinfo
  {booktitle} {{Proceedings, 14th Workshop on Gravitational wave data analysis
  (GWDAW-14): Rome, Italy, January 26-29, 2010}}},\ }\href {\doibase
  10.1088/0264-9381/27/19/194002} {\bibfield  {journal} {\bibinfo  {journal}
  {Class. Quant. Grav.}\ }\textbf {\bibinfo {volume} {27}},\ \bibinfo {pages}
  {194002} (\bibinfo {year} {2010})}\BibitemShut {NoStop}%
\bibitem [{\citenamefont {Abbott}\ \emph {et~al.}(2018)\citenamefont {Abbott}
  \emph {et~al.}}]{LIGOScientific:2018mvr}%
  \BibitemOpen
  \bibfield  {author} {\bibinfo {author} {\bibfnamefont {B.~P.}\ \bibnamefont
  {Abbott}} \emph {et~al.} (\bibinfo {collaboration} {LIGO Scientific,
  Virgo}),\ }\href@noop {} {\  (\bibinfo {year} {2018})},\ \Eprint
  {http://arxiv.org/abs/1811.12907} {arXiv:1811.12907 [astro-ph.HE]}
  \BibitemShut {NoStop}%
\bibitem [{\citenamefont {Carullo}\ \emph {et~al.}(2018)\citenamefont {Carullo}
  \emph {et~al.}}]{Carullo:2018sfu}%
  \BibitemOpen
  \bibfield  {author} {\bibinfo {author} {\bibfnamefont {G.}~\bibnamefont
  {Carullo}} \emph {et~al.},\ }\href {\doibase 10.1103/PhysRevD.98.104020}
  {\bibfield  {journal} {\bibinfo  {journal} {Phys. Rev.}\ }\textbf {\bibinfo
  {volume} {D98}},\ \bibinfo {pages} {104020} (\bibinfo {year} {2018})},\
  \Eprint {http://arxiv.org/abs/1805.04760} {arXiv:1805.04760 [gr-qc]}
  \BibitemShut {NoStop}%
\bibitem [{\citenamefont {Carullo}\ \emph {et~al.}(2019)\citenamefont
  {Carullo}, \citenamefont {Del~Pozzo},\ and\ \citenamefont
  {Veitch}}]{Carullo:2019flw}%
  \BibitemOpen
  \bibfield  {author} {\bibinfo {author} {\bibfnamefont {G.}~\bibnamefont
  {Carullo}}, \bibinfo {author} {\bibfnamefont {W.}~\bibnamefont {Del~Pozzo}},
  \ and\ \bibinfo {author} {\bibfnamefont {J.}~\bibnamefont {Veitch}},\
  }\href@noop {} {\  (\bibinfo {year} {2019})},\ \Eprint
  {http://arxiv.org/abs/1902.07527} {arXiv:1902.07527 [gr-qc]} \BibitemShut
  {NoStop}%
\bibitem [{\citenamefont {Brito}\ \emph {et~al.}(2018)\citenamefont {Brito},
  \citenamefont {Buonanno},\ and\ \citenamefont {Raymond}}]{Brito:2018rfr}%
  \BibitemOpen
  \bibfield  {author} {\bibinfo {author} {\bibfnamefont {R.}~\bibnamefont
  {Brito}}, \bibinfo {author} {\bibfnamefont {A.}~\bibnamefont {Buonanno}}, \
  and\ \bibinfo {author} {\bibfnamefont {V.}~\bibnamefont {Raymond}},\ }\href
  {\doibase 10.1103/PhysRevD.98.084038} {\bibfield  {journal} {\bibinfo
  {journal} {Phys. Rev.}\ }\textbf {\bibinfo {volume} {D98}},\ \bibinfo {pages}
  {084038} (\bibinfo {year} {2018})},\ \Eprint
  {http://arxiv.org/abs/1805.00293} {arXiv:1805.00293 [gr-qc]} \BibitemShut
  {NoStop}%
\bibitem [{\citenamefont {Isi}\ \emph {et~al.}(2019)\citenamefont {Isi},
  \citenamefont {Giesler}, \citenamefont {Farr}, \citenamefont {Scheel},\ and\
  \citenamefont {Teukolsky}}]{Isi:2019aib}%
  \BibitemOpen
  \bibfield  {author} {\bibinfo {author} {\bibfnamefont {M.}~\bibnamefont
  {Isi}}, \bibinfo {author} {\bibfnamefont {M.}~\bibnamefont {Giesler}},
  \bibinfo {author} {\bibfnamefont {W.~M.}\ \bibnamefont {Farr}}, \bibinfo
  {author} {\bibfnamefont {M.~A.}\ \bibnamefont {Scheel}}, \ and\ \bibinfo
  {author} {\bibfnamefont {S.~A.}\ \bibnamefont {Teukolsky}},\ }\href@noop {}
  {\  (\bibinfo {year} {2019})},\ \Eprint {http://arxiv.org/abs/1905.00869}
  {arXiv:1905.00869 [gr-qc]} \BibitemShut {NoStop}%
\bibitem [{\citenamefont {Luo}\ \emph {et~al.}(2016)\citenamefont {Luo} \emph
  {et~al.}}]{Luo:2015ght}%
  \BibitemOpen
  \bibfield  {author} {\bibinfo {author} {\bibfnamefont {J.}~\bibnamefont
  {Luo}} \emph {et~al.} (\bibinfo {collaboration} {TianQin}),\ }\href {\doibase
  10.1088/0264-9381/33/3/035010} {\bibfield  {journal} {\bibinfo  {journal}
  {Class. Quant. Grav.}\ }\textbf {\bibinfo {volume} {33}},\ \bibinfo {pages}
  {035010} (\bibinfo {year} {2016})},\ \Eprint
  {http://arxiv.org/abs/1512.02076} {arXiv:1512.02076 [astro-ph.IM]}
  \BibitemShut {NoStop}%
\bibitem [{\citenamefont {Wang}\ \emph {et~al.}(2019)\citenamefont {Wang} \emph
  {et~al.}}]{Wang:2019ryf}%
  \BibitemOpen
  \bibfield  {author} {\bibinfo {author} {\bibfnamefont {H.-T.}\ \bibnamefont
  {Wang}} \emph {et~al.},\ }\href@noop {} {\  (\bibinfo {year} {2019})},\
  \Eprint {http://arxiv.org/abs/1902.04423} {arXiv:1902.04423 [astro-ph.HE]}
  \BibitemShut {NoStop}%
\bibitem [{\citenamefont {Feng}\ \emph {et~al.}(2019)\citenamefont {Feng},
  \citenamefont {Wang}, \citenamefont {Hu}, \citenamefont {Hu},\ and\
  \citenamefont {Wang}}]{Feng:2019wgq}%
  \BibitemOpen
  \bibfield  {author} {\bibinfo {author} {\bibfnamefont {W.-F.}\ \bibnamefont
  {Feng}}, \bibinfo {author} {\bibfnamefont {H.-T.}\ \bibnamefont {Wang}},
  \bibinfo {author} {\bibfnamefont {X.-C.}\ \bibnamefont {Hu}}, \bibinfo
  {author} {\bibfnamefont {Y.-M.}\ \bibnamefont {Hu}}, \ and\ \bibinfo {author}
  {\bibfnamefont {Y.}~\bibnamefont {Wang}},\ }\href@noop {} {\  (\bibinfo
  {year} {2019})},\ \Eprint {http://arxiv.org/abs/1901.02159} {arXiv:1901.02159
  [astro-ph.IM]} \BibitemShut {NoStop}%
\bibitem [{\citenamefont {Hu}\ \emph {et~al.}(2017)\citenamefont {Hu},
  \citenamefont {Mei},\ and\ \citenamefont {Luo}}]{Hu:2017yoc}%
  \BibitemOpen
  \bibfield  {author} {\bibinfo {author} {\bibfnamefont {Y.-M.}\ \bibnamefont
  {Hu}}, \bibinfo {author} {\bibfnamefont {J.}~\bibnamefont {Mei}}, \ and\
  \bibinfo {author} {\bibfnamefont {J.}~\bibnamefont {Luo}},\ }\href {\doibase
  10.1093/nsr/nwx115} {\bibfield  {journal} {\bibinfo  {journal} {Natl. Sci.
  Rev.}\ }\textbf {\bibinfo {volume} {4}},\ \bibinfo {pages} {683} (\bibinfo
  {year} {2017})}\BibitemShut {NoStop}%
\bibitem [{\citenamefont {Audley}\ \emph {et~al.}(2017)\citenamefont {Audley}
  \emph {et~al.}}]{Audley:2017drz}%
  \BibitemOpen
  \bibfield  {author} {\bibinfo {author} {\bibfnamefont {H.}~\bibnamefont
  {Audley}} \emph {et~al.} (\bibinfo {collaboration} {LISA}),\ }\href@noop {}
  {\  (\bibinfo {year} {2017})},\ \Eprint {http://arxiv.org/abs/1702.00786}
  {arXiv:1702.00786 [astro-ph.IM]} \BibitemShut {NoStop}%
\bibitem [{\citenamefont {Chandrasekhar}(1984)}]{Chandrasekhar:1984siy}%
  \BibitemOpen
  \bibfield  {author} {\bibinfo {author} {\bibfnamefont {S.}~\bibnamefont
  {Chandrasekhar}},\ }\bibfield  {booktitle} {\emph {\bibinfo {booktitle}
  {{Proceedings, 10th International Conference on General Relativity and
  Gravitation: Padua, Italy, July 4-9, 1983}}},\ }\href {\doibase
  10.1007/978-94-009-6469-3_2} {\bibfield  {journal} {\bibinfo  {journal}
  {Fundam. Theor. Phys.}\ }\textbf {\bibinfo {volume} {9}},\ \bibinfo {pages}
  {5} (\bibinfo {year} {1984})}\BibitemShut {NoStop}%
\bibitem [{\citenamefont {Chandrasekhar}\ and\ \citenamefont
  {Detweiler}(1975)}]{Chandrasekhar:1975zza}%
  \BibitemOpen
  \bibfield  {author} {\bibinfo {author} {\bibfnamefont {S.}~\bibnamefont
  {Chandrasekhar}}\ and\ \bibinfo {author} {\bibfnamefont {S.~L.}\ \bibnamefont
  {Detweiler}},\ }\href {\doibase 10.1098/rspa.1975.0112} {\bibfield  {journal}
  {\bibinfo  {journal} {Proc. Roy. Soc. Lond.}\ }\textbf {\bibinfo {volume}
  {A344}},\ \bibinfo {pages} {441} (\bibinfo {year} {1975})}\BibitemShut
  {NoStop}%
\bibitem [{\citenamefont {Leaver}(1985)}]{Leaver:1985ax}%
  \BibitemOpen
  \bibfield  {author} {\bibinfo {author} {\bibfnamefont {E.~W.}\ \bibnamefont
  {Leaver}},\ }\href {\doibase 10.1098/rspa.1985.0119} {\bibfield  {journal}
  {\bibinfo  {journal} {Proc. Roy. Soc. Lond.}\ }\textbf {\bibinfo {volume}
  {A402}},\ \bibinfo {pages} {285} (\bibinfo {year} {1985})}\BibitemShut
  {NoStop}%
\bibitem [{\citenamefont {Onozawa}(1997)}]{Onozawa:1996ux}%
  \BibitemOpen
  \bibfield  {author} {\bibinfo {author} {\bibfnamefont {H.}~\bibnamefont
  {Onozawa}},\ }\href {\doibase 10.1103/PhysRevD.55.3593} {\bibfield  {journal}
  {\bibinfo  {journal} {Phys. Rev.}\ }\textbf {\bibinfo {volume} {D55}},\
  \bibinfo {pages} {3593} (\bibinfo {year} {1997})},\ \Eprint
  {http://arxiv.org/abs/gr-qc/9610048} {arXiv:gr-qc/9610048 [gr-qc]}
  \BibitemShut {NoStop}%
\bibitem [{\citenamefont {Berti}\ \emph {et~al.}(2003)\citenamefont {Berti},
  \citenamefont {Cardoso}, \citenamefont {Kokkotas},\ and\ \citenamefont
  {Onozawa}}]{Berti:2003jh}%
  \BibitemOpen
  \bibfield  {author} {\bibinfo {author} {\bibfnamefont {E.}~\bibnamefont
  {Berti}}, \bibinfo {author} {\bibfnamefont {V.}~\bibnamefont {Cardoso}},
  \bibinfo {author} {\bibfnamefont {K.~D.}\ \bibnamefont {Kokkotas}}, \ and\
  \bibinfo {author} {\bibfnamefont {H.}~\bibnamefont {Onozawa}},\ }\href
  {\doibase 10.1103/PhysRevD.68.124018} {\bibfield  {journal} {\bibinfo
  {journal} {Phys. Rev.}\ }\textbf {\bibinfo {volume} {D68}},\ \bibinfo {pages}
  {124018} (\bibinfo {year} {2003})},\ \Eprint
  {http://arxiv.org/abs/hep-th/0307013} {arXiv:hep-th/0307013 [hep-th]}
  \BibitemShut {NoStop}%
\bibitem [{\citenamefont {Berti}\ and\ \citenamefont
  {Kokkotas}(2005)}]{Berti:2005eb}%
  \BibitemOpen
  \bibfield  {author} {\bibinfo {author} {\bibfnamefont {E.}~\bibnamefont
  {Berti}}\ and\ \bibinfo {author} {\bibfnamefont {K.~D.}\ \bibnamefont
  {Kokkotas}},\ }\href {\doibase 10.1103/PhysRevD.71.124008} {\bibfield
  {journal} {\bibinfo  {journal} {Phys. Rev.}\ }\textbf {\bibinfo {volume}
  {D71}},\ \bibinfo {pages} {124008} (\bibinfo {year} {2005})},\ \Eprint
  {http://arxiv.org/abs/gr-qc/0502065} {arXiv:gr-qc/0502065 [gr-qc]}
  \BibitemShut {NoStop}%
\bibitem [{\citenamefont {Kokkotas}(1993)}]{Kokkotas:1993ef}%
  \BibitemOpen
  \bibfield  {author} {\bibinfo {author} {\bibfnamefont {K.~D.}\ \bibnamefont
  {Kokkotas}},\ }\href {\doibase 10.1007/BF02822861} {\bibfield  {journal}
  {\bibinfo  {journal} {Nuovo Cim.}\ }\textbf {\bibinfo {volume} {B108}},\
  \bibinfo {pages} {991} (\bibinfo {year} {1993})}\BibitemShut {NoStop}%
\bibitem [{\citenamefont {Meidam}\ \emph {et~al.}(2014)\citenamefont {Meidam},
  \citenamefont {Agathos}, \citenamefont {Van Den~Broeck}, \citenamefont
  {Veitch},\ and\ \citenamefont {Sathyaprakash}}]{Meidam:2014jpa}%
  \BibitemOpen
  \bibfield  {author} {\bibinfo {author} {\bibfnamefont {J.}~\bibnamefont
  {Meidam}}, \bibinfo {author} {\bibfnamefont {M.}~\bibnamefont {Agathos}},
  \bibinfo {author} {\bibfnamefont {C.}~\bibnamefont {Van Den~Broeck}},
  \bibinfo {author} {\bibfnamefont {J.}~\bibnamefont {Veitch}}, \ and\ \bibinfo
  {author} {\bibfnamefont {B.~S.}\ \bibnamefont {Sathyaprakash}},\ }\href
  {\doibase 10.1103/PhysRevD.90.064009} {\bibfield  {journal} {\bibinfo
  {journal} {Phys. Rev.}\ }\textbf {\bibinfo {volume} {D90}},\ \bibinfo {pages}
  {064009} (\bibinfo {year} {2014})},\ \Eprint {http://arxiv.org/abs/1406.3201}
  {arXiv:1406.3201 [gr-qc]} \BibitemShut {NoStop}%
\bibitem [{\citenamefont {London}\ \emph {et~al.}(2014)\citenamefont {London},
  \citenamefont {Shoemaker},\ and\ \citenamefont {Healy}}]{London:2014cma}%
  \BibitemOpen
  \bibfield  {author} {\bibinfo {author} {\bibfnamefont {L.}~\bibnamefont
  {London}}, \bibinfo {author} {\bibfnamefont {D.}~\bibnamefont {Shoemaker}}, \
  and\ \bibinfo {author} {\bibfnamefont {J.}~\bibnamefont {Healy}},\ }\href
  {\doibase 10.1103/PhysRevD.90.124032, 10.1103/PhysRevD.94.069902} {\bibfield
  {journal} {\bibinfo  {journal} {Phys. Rev.}\ }\textbf {\bibinfo {volume}
  {D90}},\ \bibinfo {pages} {124032} (\bibinfo {year} {2014})},\ \bibinfo
  {note} {[Erratum: Phys. Rev.D94,no.6,069902(2016)]},\ \Eprint
  {http://arxiv.org/abs/1404.3197} {arXiv:1404.3197 [gr-qc]} \BibitemShut
  {NoStop}%
\bibitem [{\citenamefont {London}(2018)}]{London:2018gaq}%
  \BibitemOpen
  \bibfield  {author} {\bibinfo {author} {\bibfnamefont {L.~T.}\ \bibnamefont
  {London}},\ }\href@noop {} {\  (\bibinfo {year} {2018})},\ \Eprint
  {http://arxiv.org/abs/1801.08208} {arXiv:1801.08208 [gr-qc]} \BibitemShut
  {NoStop}%
\bibitem [{\citenamefont {Husa}\ \emph {et~al.}(2016)\citenamefont {Husa},
  \citenamefont {Khan}, \citenamefont {Hannam}, \citenamefont {P��rrer},
  \citenamefont {Ohme}, \citenamefont {Jim��nez~Forteza},\ and\
  \citenamefont {Boh��}}]{Husa:2015iqa}%
  \BibitemOpen
  \bibfield  {author} {\bibinfo {author} {\bibfnamefont {S.}~\bibnamefont
  {Husa}}, \bibinfo {author} {\bibfnamefont {S.}~\bibnamefont {Khan}}, \bibinfo
  {author} {\bibfnamefont {M.}~\bibnamefont {Hannam}}, \bibinfo {author}
  {\bibfnamefont {M.}~\bibnamefont {P��rrer}}, \bibinfo {author}
  {\bibfnamefont {F.}~\bibnamefont {Ohme}}, \bibinfo {author} {\bibfnamefont
  {X.}~\bibnamefont {Jim��nez~Forteza}}, \ and\ \bibinfo {author}
  {\bibfnamefont {A.}~\bibnamefont {Boh��}},\ }\href {\doibase
  10.1103/PhysRevD.93.044006} {\bibfield  {journal} {\bibinfo  {journal} {Phys.
  Rev.}\ }\textbf {\bibinfo {volume} {D93}},\ \bibinfo {pages} {044006}
  (\bibinfo {year} {2016})},\ \Eprint {http://arxiv.org/abs/1508.07250}
  {arXiv:1508.07250 [gr-qc]} \BibitemShut {NoStop}%
\bibitem [{\citenamefont {Hofmann}\ \emph {et~al.}(2016)\citenamefont
  {Hofmann}, \citenamefont {Barausse},\ and\ \citenamefont
  {Rezzolla}}]{Hofmann:2016yih}%
  \BibitemOpen
  \bibfield  {author} {\bibinfo {author} {\bibfnamefont {F.}~\bibnamefont
  {Hofmann}}, \bibinfo {author} {\bibfnamefont {E.}~\bibnamefont {Barausse}}, \
  and\ \bibinfo {author} {\bibfnamefont {L.}~\bibnamefont {Rezzolla}},\ }\href
  {\doibase 10.3847/2041-8205/825/2/L19} {\bibfield  {journal} {\bibinfo
  {journal} {Astrophys. J.}\ }\textbf {\bibinfo {volume} {825}},\ \bibinfo
  {pages} {L19} (\bibinfo {year} {2016})},\ \Eprint
  {http://arxiv.org/abs/1605.01938} {arXiv:1605.01938 [gr-qc]} \BibitemShut
  {NoStop}%
\bibitem [{\citenamefont {Barausse}\ \emph {et~al.}(2012)\citenamefont
  {Barausse}, \citenamefont {Morozova},\ and\ \citenamefont
  {Rezzolla}}]{Barausse:2012qz}%
  \BibitemOpen
  \bibfield  {author} {\bibinfo {author} {\bibfnamefont {E.}~\bibnamefont
  {Barausse}}, \bibinfo {author} {\bibfnamefont {V.}~\bibnamefont {Morozova}},
  \ and\ \bibinfo {author} {\bibfnamefont {L.}~\bibnamefont {Rezzolla}},\
  }\href {\doibase 10.1088/0004-637X/786/1/76, 10.1088/0004-637X/758/1/63}
  {\bibfield  {journal} {\bibinfo  {journal} {Astrophys. J.}\ }\textbf
  {\bibinfo {volume} {758}},\ \bibinfo {pages} {63} (\bibinfo {year} {2012})},\
  \bibinfo {note} {[Erratum: Astrophys. J.786,76(2014)]},\ \Eprint
  {http://arxiv.org/abs/1206.3803} {arXiv:1206.3803 [gr-qc]} \BibitemShut
  {NoStop}%
\bibitem [{\citenamefont {Li}\ \emph {et~al.}(2012)\citenamefont {Li},
  \citenamefont {Del~Pozzo}, \citenamefont {Vitale}, \citenamefont {Van
  Den~Broeck}, \citenamefont {Agathos}, \citenamefont {Veitch}, \citenamefont
  {Grover}, \citenamefont {Sidery}, \citenamefont {Sturani},\ and\
  \citenamefont {Vecchio}}]{Li:2011cg}%
  \BibitemOpen
  \bibfield  {author} {\bibinfo {author} {\bibfnamefont {T.~G.~F.}\
  \bibnamefont {Li}}, \bibinfo {author} {\bibfnamefont {W.}~\bibnamefont
  {Del~Pozzo}}, \bibinfo {author} {\bibfnamefont {S.}~\bibnamefont {Vitale}},
  \bibinfo {author} {\bibfnamefont {C.}~\bibnamefont {Van Den~Broeck}},
  \bibinfo {author} {\bibfnamefont {M.}~\bibnamefont {Agathos}}, \bibinfo
  {author} {\bibfnamefont {J.}~\bibnamefont {Veitch}}, \bibinfo {author}
  {\bibfnamefont {K.}~\bibnamefont {Grover}}, \bibinfo {author} {\bibfnamefont
  {T.}~\bibnamefont {Sidery}}, \bibinfo {author} {\bibfnamefont
  {R.}~\bibnamefont {Sturani}}, \ and\ \bibinfo {author} {\bibfnamefont
  {A.}~\bibnamefont {Vecchio}},\ }\href {\doibase 10.1103/PhysRevD.85.082003}
  {\bibfield  {journal} {\bibinfo  {journal} {Phys. Rev.}\ }\textbf {\bibinfo
  {volume} {D85}},\ \bibinfo {pages} {082003} (\bibinfo {year} {2012})},\
  \Eprint {http://arxiv.org/abs/1110.0530} {arXiv:1110.0530 [gr-qc]}
  \BibitemShut {NoStop}%
\bibitem [{\citenamefont {Hu}\ \emph {et~al.}(2018)\citenamefont {Hu},
  \citenamefont {Li}, \citenamefont {Wang}, \citenamefont {Feng}, \citenamefont
  {Zhou}, \citenamefont {Hu}, \citenamefont {Hu}, \citenamefont {Mei},\ and\
  \citenamefont {Shao}}]{Hu:2018yqb}%
  \BibitemOpen
  \bibfield  {author} {\bibinfo {author} {\bibfnamefont {X.-C.}\ \bibnamefont
  {Hu}}, \bibinfo {author} {\bibfnamefont {X.-H.}\ \bibnamefont {Li}}, \bibinfo
  {author} {\bibfnamefont {Y.}~\bibnamefont {Wang}}, \bibinfo {author}
  {\bibfnamefont {W.-F.}\ \bibnamefont {Feng}}, \bibinfo {author}
  {\bibfnamefont {M.-Y.}\ \bibnamefont {Zhou}}, \bibinfo {author}
  {\bibfnamefont {Y.-M.}\ \bibnamefont {Hu}}, \bibinfo {author} {\bibfnamefont
  {S.-C.}\ \bibnamefont {Hu}}, \bibinfo {author} {\bibfnamefont {J.-W.}\
  \bibnamefont {Mei}}, \ and\ \bibinfo {author} {\bibfnamefont {C.-G.}\
  \bibnamefont {Shao}},\ }\href {\doibase 10.1088/1361-6382/aab52f} {\bibfield
  {journal} {\bibinfo  {journal} {Class. Quant. Grav.}\ }\textbf {\bibinfo
  {volume} {35}},\ \bibinfo {pages} {095008} (\bibinfo {year} {2018})},\
  \Eprint {http://arxiv.org/abs/1803.03368} {arXiv:1803.03368 [gr-qc]}
  \BibitemShut {NoStop}%
\bibitem [{\citenamefont {Robson}\ \emph {et~al.}(2019)\citenamefont {Robson},
  \citenamefont {Cornish},\ and\ \citenamefont {Liu}}]{Cornish:2018dyw}%
  \BibitemOpen
  \bibfield  {author} {\bibinfo {author} {\bibfnamefont {T.}~\bibnamefont
  {Robson}}, \bibinfo {author} {\bibfnamefont {N.}~\bibnamefont {Cornish}}, \
  and\ \bibinfo {author} {\bibfnamefont {C.}~\bibnamefont {Liu}},\ }\href
  {\doibase 10.1088/1361-6382/ab1101} {\bibfield  {journal} {\bibinfo
  {journal} {Class. Quant. Grav.}\ }\textbf {\bibinfo {volume} {36}},\ \bibinfo
  {pages} {105011} (\bibinfo {year} {2019})},\ \Eprint
  {http://arxiv.org/abs/1803.01944} {arXiv:1803.01944 [astro-ph.HE]}
  \BibitemShut {NoStop}%
\bibitem [{\citenamefont {Finn}(1992)}]{Finn:1992wt}%
  \BibitemOpen
  \bibfield  {author} {\bibinfo {author} {\bibfnamefont {L.~S.}\ \bibnamefont
  {Finn}},\ }\href {\doibase 10.1103/PhysRevD.46.5236} {\bibfield  {journal}
  {\bibinfo  {journal} {Phys. Rev.}\ }\textbf {\bibinfo {volume} {D46}},\
  \bibinfo {pages} {5236} (\bibinfo {year} {1992})},\ \Eprint
  {http://arxiv.org/abs/gr-qc/9209010} {arXiv:gr-qc/9209010 [gr-qc]}
  \BibitemShut {NoStop}%
\bibitem [{\citenamefont {Baibhav}\ \emph {et~al.}(2018)\citenamefont
  {Baibhav}, \citenamefont {Berti}, \citenamefont {Cardoso},\ and\
  \citenamefont {Khanna}}]{Baibhav:2017jhs}%
  \BibitemOpen
  \bibfield  {author} {\bibinfo {author} {\bibfnamefont {V.}~\bibnamefont
  {Baibhav}}, \bibinfo {author} {\bibfnamefont {E.}~\bibnamefont {Berti}},
  \bibinfo {author} {\bibfnamefont {V.}~\bibnamefont {Cardoso}}, \ and\
  \bibinfo {author} {\bibfnamefont {G.}~\bibnamefont {Khanna}},\ }\href
  {\doibase 10.1103/PhysRevD.97.044048} {\bibfield  {journal} {\bibinfo
  {journal} {Phys. Rev.}\ }\textbf {\bibinfo {volume} {D97}},\ \bibinfo {pages}
  {044048} (\bibinfo {year} {2018})},\ \Eprint
  {http://arxiv.org/abs/1710.02156} {arXiv:1710.02156 [gr-qc]} \BibitemShut
  {NoStop}%
\bibitem [{\citenamefont {Cutler}\ and\ \citenamefont
  {Flanagan}(1994)}]{Cutler:1994ys}%
  \BibitemOpen
  \bibfield  {author} {\bibinfo {author} {\bibfnamefont {C.}~\bibnamefont
  {Cutler}}\ and\ \bibinfo {author} {\bibfnamefont {E.~E.}\ \bibnamefont
  {Flanagan}},\ }\href {\doibase 10.1103/PhysRevD.49.2658} {\bibfield
  {journal} {\bibinfo  {journal} {Phys. Rev.}\ }\textbf {\bibinfo {volume}
  {D49}},\ \bibinfo {pages} {2658} (\bibinfo {year} {1994})},\ \Eprint
  {http://arxiv.org/abs/gr-qc/9402014} {arXiv:gr-qc/9402014 [gr-qc]}
  \BibitemShut {NoStop}%
\bibitem [{\citenamefont {Vallisneri}(2008)}]{Vallisneri:2007ev}%
  \BibitemOpen
  \bibfield  {author} {\bibinfo {author} {\bibfnamefont {M.}~\bibnamefont
  {Vallisneri}},\ }\href {\doibase 10.1103/PhysRevD.77.042001} {\bibfield
  {journal} {\bibinfo  {journal} {Phys. Rev.}\ }\textbf {\bibinfo {volume}
  {D77}},\ \bibinfo {pages} {042001} (\bibinfo {year} {2008})},\ \Eprint
  {http://arxiv.org/abs/gr-qc/0703086} {arXiv:gr-qc/0703086 [GR-QC]}
  \BibitemShut {NoStop}%
\bibitem [{\citenamefont {{Klein}}\ \emph {et~al.}(2016)\citenamefont
  {{Klein}}, \citenamefont {{Barausse}}, \citenamefont {{Sesana}},
  \citenamefont {{Petiteau}}, \citenamefont {{Berti}}, \citenamefont {{Babak}},
  \citenamefont {{Gair}}, \citenamefont {{Aoudia}}, \citenamefont {{Hinder}},
  \citenamefont {{Ohme}},\ and\ \citenamefont {{Wardell}}}]{Klein16}%
  \BibitemOpen
  \bibfield  {author} {\bibinfo {author} {\bibfnamefont {A.}~\bibnamefont
  {{Klein}}}, \bibinfo {author} {\bibfnamefont {E.}~\bibnamefont {{Barausse}}},
  \bibinfo {author} {\bibfnamefont {A.}~\bibnamefont {{Sesana}}}, \bibinfo
  {author} {\bibfnamefont {A.}~\bibnamefont {{Petiteau}}}, \bibinfo {author}
  {\bibfnamefont {E.}~\bibnamefont {{Berti}}}, \bibinfo {author} {\bibfnamefont
  {S.}~\bibnamefont {{Babak}}}, \bibinfo {author} {\bibfnamefont
  {J.}~\bibnamefont {{Gair}}}, \bibinfo {author} {\bibfnamefont
  {S.}~\bibnamefont {{Aoudia}}}, \bibinfo {author} {\bibfnamefont
  {I.}~\bibnamefont {{Hinder}}}, \bibinfo {author} {\bibfnamefont
  {F.}~\bibnamefont {{Ohme}}}, \ and\ \bibinfo {author} {\bibfnamefont
  {B.}~\bibnamefont {{Wardell}}},\ }\href {\doibase 10.1103/PhysRevD.93.024003}
  {\bibfield  {journal} {\bibinfo  {journal} {\prd}\ }\textbf {\bibinfo
  {volume} {93}},\ \bibinfo {eid} {024003} (\bibinfo {year} {2016})},\ \Eprint
  {http://arxiv.org/abs/1511.05581} {arXiv:1511.05581 [gr-qc]} \BibitemShut
  {NoStop}%
\bibitem [{\citenamefont {{Barausse}}(2012)}]{EB12}%
  \BibitemOpen
  \bibfield  {author} {\bibinfo {author} {\bibfnamefont {E.}~\bibnamefont
  {{Barausse}}},\ }\href {\doibase 10.1111/j.1365-2966.2012.21057.x} {\bibfield
   {journal} {\bibinfo  {journal} {Monthly Notices of the Royal Astronomical
  Society}\ }\textbf {\bibinfo {volume} {423}},\ \bibinfo {pages} {2533}
  (\bibinfo {year} {2012})},\ \Eprint {http://arxiv.org/abs/1201.5888}
  {arXiv:1201.5888} \BibitemShut {NoStop}%
\bibitem [{\citenamefont {{Sesana}}\ \emph {et~al.}(2014)\citenamefont
  {{Sesana}}, \citenamefont {{Barausse}}, \citenamefont {{Dotti}},\ and\
  \citenamefont {{Rossi}}}]{Sesana14}%
  \BibitemOpen
  \bibfield  {author} {\bibinfo {author} {\bibfnamefont {A.}~\bibnamefont
  {{Sesana}}}, \bibinfo {author} {\bibfnamefont {E.}~\bibnamefont
  {{Barausse}}}, \bibinfo {author} {\bibfnamefont {M.}~\bibnamefont {{Dotti}}},
  \ and\ \bibinfo {author} {\bibfnamefont {E.~M.}\ \bibnamefont {{Rossi}}},\
  }\href {\doibase 10.1088/0004-637X/794/2/104} {\bibfield  {journal} {\bibinfo
   {journal} {APJ}\ }\textbf {\bibinfo {volume} {794}},\ \bibinfo {eid} {104}
  (\bibinfo {year} {2014})},\ \Eprint {http://arxiv.org/abs/1402.7088}
  {arXiv:1402.7088} \BibitemShut {NoStop}%
\bibitem [{\citenamefont {{Antonini}}\ \emph {et~al.}(2015)\citenamefont
  {{Antonini}}, \citenamefont {{Barausse}},\ and\ \citenamefont
  {{Silk}}}]{Antonini_long}%
  \BibitemOpen
  \bibfield  {author} {\bibinfo {author} {\bibfnamefont {F.}~\bibnamefont
  {{Antonini}}}, \bibinfo {author} {\bibfnamefont {E.}~\bibnamefont
  {{Barausse}}}, \ and\ \bibinfo {author} {\bibfnamefont {J.}~\bibnamefont
  {{Silk}}},\ }\href {\doibase 10.1088/0004-637X/812/1/72} {\bibfield
  {journal} {\bibinfo  {journal} {APJ}\ }\textbf {\bibinfo {volume} {812}},\
  \bibinfo {eid} {72} (\bibinfo {year} {2015})},\ \Eprint
  {http://arxiv.org/abs/1506.02050} {arXiv:1506.02050} \BibitemShut {NoStop}%
\bibitem [{\citenamefont {Madau}\ and\ \citenamefont
  {Rees}(2001)}]{Madau:2001sc}%
  \BibitemOpen
  \bibfield  {author} {\bibinfo {author} {\bibfnamefont {P.}~\bibnamefont
  {Madau}}\ and\ \bibinfo {author} {\bibfnamefont {M.~J.}\ \bibnamefont
  {Rees}},\ }\href {\doibase 10.1086/319848} {\bibfield  {journal} {\bibinfo
  {journal} {Astrophys. J.}\ }\textbf {\bibinfo {volume} {551}},\ \bibinfo
  {pages} {L27} (\bibinfo {year} {2001})},\ \Eprint
  {http://arxiv.org/abs/astro-ph/0101223} {arXiv:astro-ph/0101223 [astro-ph]}
  \BibitemShut {NoStop}%
\bibitem [{\citenamefont {Bromm}\ and\ \citenamefont
  {Loeb}(2003)}]{Bromm:2002hb}%
  \BibitemOpen
  \bibfield  {author} {\bibinfo {author} {\bibfnamefont {V.}~\bibnamefont
  {Bromm}}\ and\ \bibinfo {author} {\bibfnamefont {A.}~\bibnamefont {Loeb}},\
  }\href {\doibase 10.1086/377529} {\bibfield  {journal} {\bibinfo  {journal}
  {Astrophys. J.}\ }\textbf {\bibinfo {volume} {596}},\ \bibinfo {pages} {34}
  (\bibinfo {year} {2003})},\ \Eprint {http://arxiv.org/abs/astro-ph/0212400}
  {arXiv:astro-ph/0212400 [astro-ph]} \BibitemShut {NoStop}%
\bibitem [{\citenamefont {Begelman}\ \emph {et~al.}(2006)\citenamefont
  {Begelman}, \citenamefont {Volonteri},\ and\ \citenamefont
  {Rees}}]{Begelman:2006db}%
  \BibitemOpen
  \bibfield  {author} {\bibinfo {author} {\bibfnamefont {M.~C.}\ \bibnamefont
  {Begelman}}, \bibinfo {author} {\bibfnamefont {M.}~\bibnamefont {Volonteri}},
  \ and\ \bibinfo {author} {\bibfnamefont {M.~J.}\ \bibnamefont {Rees}},\
  }\href {\doibase 10.1111/j.1365-2966.2006.10467.x} {\bibfield  {journal}
  {\bibinfo  {journal} {Mon. Not. Roy. Astron. Soc.}\ }\textbf {\bibinfo
  {volume} {370}},\ \bibinfo {pages} {289} (\bibinfo {year} {2006})},\ \Eprint
  {http://arxiv.org/abs/astro-ph/0602363} {arXiv:astro-ph/0602363 [astro-ph]}
  \BibitemShut {NoStop}%
\bibitem [{\citenamefont {Lodato}\ and\ \citenamefont
  {Natarajan}(2006)}]{Lodato:2006hw}%
  \BibitemOpen
  \bibfield  {author} {\bibinfo {author} {\bibfnamefont {G.}~\bibnamefont
  {Lodato}}\ and\ \bibinfo {author} {\bibfnamefont {P.}~\bibnamefont
  {Natarajan}},\ }\href {\doibase 10.1111/j.1365-2966.2006.10801.x} {\bibfield
  {journal} {\bibinfo  {journal} {Mon. Not. Roy. Astron. Soc.}\ }\textbf
  {\bibinfo {volume} {371}},\ \bibinfo {pages} {1813} (\bibinfo {year}
  {2006})},\ \Eprint {http://arxiv.org/abs/astro-ph/0606159}
  {arXiv:astro-ph/0606159 [astro-ph]} \BibitemShut {NoStop}%
\bibitem [{\citenamefont {Bao}\ \emph {et~al.}(2019)\citenamefont {Bao},
  \citenamefont {Shi}, \citenamefont {Wang}, \citenamefont {Zhang},
  \citenamefont {Hu}, \citenamefont {Mei},\ and\ \citenamefont
  {Luo}}]{Bao:2019kgt}%
  \BibitemOpen
  \bibfield  {author} {\bibinfo {author} {\bibfnamefont {J.}~\bibnamefont
  {Bao}}, \bibinfo {author} {\bibfnamefont {C.}~\bibnamefont {Shi}}, \bibinfo
  {author} {\bibfnamefont {H.}~\bibnamefont {Wang}}, \bibinfo {author}
  {\bibfnamefont {J.-d.}\ \bibnamefont {Zhang}}, \bibinfo {author}
  {\bibfnamefont {Y.}~\bibnamefont {Hu}}, \bibinfo {author} {\bibfnamefont
  {J.}~\bibnamefont {Mei}}, \ and\ \bibinfo {author} {\bibfnamefont
  {J.}~\bibnamefont {Luo}},\ }\href@noop {} {\  (\bibinfo {year} {2019})},\
  \Eprint {http://arxiv.org/abs/1905.11674} {arXiv:1905.11674 [gr-qc]}
  \BibitemShut {NoStop}%
\end{thebibliography}%
\end{document}